\DocumentMetadata{}
\documentclass[nonacm]{acmart}
\usepackage{subfig}
\usepackage{graphicx}
\usepackage{color,soul}
\usepackage{amsmath}
\usepackage{xspace}
\usepackage{tikz}

\usepackage{newtxtext,newtxmath}
\usepackage{hyperref}

\newcommand*\circled[2]{\tikz[baseline=(char.base)]{
            \node[shape=circle,fill=black,inner sep=1pt] (char) {\textcolor{#1}{{\footnotesize #2}}};}}

\ifx\figurename\undefined \def\figurename{Figure}\fi
\renewcommand{\figurename}{Fig.}
\renewcommand{\paragraph}[1]{\textbf{#1} }

\newcommand{\Sect}[1]{Sec.~\ref{#1}}
\newcommand{\Fig}[1]{Fig.~\ref{#1}}

\newcommand{\proj}{\textsc{Potamoi}\xspace}
\newcommand{\algo}{\textsc{SpaRW}\xspace}
\newcommand{\mode}[1]{\underline{\textsc{#1}}\xspace}

\newcommand{\no}[1]{#1}
\renewcommand{\no}[1]{#1}
\newcommand{\RNum}[1]{\uppercase\expandafter{\romannumeral #1\relax}}

\def\cG{{\mathcal{G}}}
\def\cF{{\mathcal{F}}}
\def\cI{{\mathcal{I}}}

\graphicspath{{figs/}}
\begin{document}

\title{\proj: Accelerating Neural Rendering via a Unified Streaming Architecture}

\author{Yu Feng}
\authornote{Equal contribution.}
\email{y-feng@sjtu.edu.cn}
\affiliation{%
  \institution{Shanghai Jiao Tong University}
  \city{Shanghai}
  \country{China}
}

\author{Weikai Lin}
\authornotemark[1]
\affiliation{%
  \institution{University of Rochester}
  \city{Rochester}
  \country{United States}}
\email{wlin33@ur.rochester.edu}

\author{Zihan Liu}
\email{altair.liu@sjtu.edu.cn}
\affiliation{%
  \institution{Shanghai Jiao Tong University}
  \city{Shanghai}
  \country{China}
}

\author{Jingwen Leng}
\authornote{Corresponding authors.}
\email{leng-jw@cs.sjtu.edu.cn}
\author{Minyi Guo}
\email{guo-my@cs.sjtu.edu.cn}
\affiliation{%
  \institution{Shanghai Jiao Tong University, Shanghai Qi Zhi Institute}
  \city{Shanghai}
  \country{China}
}

\author{Han Zhao}
\email{zhao-han@cs.sjtu.edu.cn}

\author{Xiaofeng Hou}
\email{hou-xf@cs.sjtu.edu.cn}

\author{Jieru Zhao}
\affiliation{%
  \institution{Shanghai Jiao Tong University}
  \city{Shanghai}
  \country{China}
}
\email{zhao-jieru@sjtu.edu.cn}

\author{Yuhao Zhu}
\authornotemark[2]
\affiliation{%
  \institution{University of Rochester}
  \city{Rochester}
  \country{United States}}
\email{yzhu@rochester.edu}

\renewcommand{\shortauthors}{Feng et al.}

\begin{abstract}

Neural Radiance Field (NeRF) has emerged as a promising alternative for photorealistic rendering. 
Despite recent algorithmic advancements, achieving real-time performance on today's resource-constrained devices remains challenging.
In this paper, we identify the primary bottlenecks in current NeRF algorithms and introduce a unified algorithm-architecture co-design, \proj, designed to accommodate various NeRF algorithms. 
Specifically, we introduce a runtime system featuring a plug-and-play algorithm, \algo, which significantly reduces the per-frame computational workload and alleviates compute inefficiencies.
Furthermore, our unified streaming pipeline coupled with customized hardware support effectively tames both SRAM and DRAM inefficiencies by minimizing repetitive DRAM access and completely eliminating SRAM bank conflicts. 
When evaluated against a baseline utilizing a dedicated DNN accelerator, our framework demonstrates a speed-up and energy reduction of 53.1$\times$ and 67.7$\times$, respectively, all while maintaining high visual quality with less than a 1.0 dB reduction in peak signal-to-noise ratio.

\end{abstract}

\maketitle

\section{Introduction}
\label{sec:intro}


Neural Radiance Field (NeRF) model~\cite{mildenhall2021nerf} revitalizes classic image-based rendering techniques~\cite{shum2008image, szeliski2022image} using modern deep learning approaches, offering a compelling alternative to conventional photorealistic rendering methods like path tracing~\cite{pharr2023physically, deng2017toward, pantaleoni2010hlbvh}.
However, the original NeRF algorithm is notoriously slow~\cite{mildenhall2021nerf}, which has spurred numerous efforts to reduce the computational demands of NeRF rendering~\cite{chen2022tensorf, sun2022direct, olszewski2023hashcc, yu2021plenoctrees, muller2022instant, hu2022efficientnerf}. 
Despite these efforts, achieving real-time performance on mobile devices remains unsolved. For example, on a mobile Volta GPU in Nvidia's Xavier SoC, models like DirectVoxGO~\cite{sun2022direct} can only achieve 0.8 Frames Per Second (FPS). Even the latest developments, 3D Gaussian Splatting (3DGS)~\cite{kerbl20233d}, can barely reach 5 FPS for rendering an $800 \times 800$ frame on Xavier.


Although accelerating NeRF is critical, it is crucial not to over-specialize hardware for any specific model due to the fast evolution in algorithm design. 
Since its introduction, NeRF models have undergone significant algorithmic transformations, evolving from the original Multilayer Perceptron (MLP)-based designs~\cite{mildenhall2021nerf} to those incorporating structured representations~\cite{yu2021plenoctrees, muller2022instant, sun2022direct}, and most recently, to implementations utilizing 3DGS with unstructured representations~\cite{kerbl20233d, lee2023compact, wu2024recent, chen2024survey}. 
Given the ongoing NeRF development, it is conceivable that NeRF models will continue to evolve. 
Therefore, system and architectural support should address the fundamental bottlenecks inherent to NeRF models as a whole, rather than optimizing for artifact features specific to individual models.


This paper introduces \proj, an algorithm-architecture co-designed approach to address fundamental inefficiencies in both computation and memory accesses inherent to a diverse set of NeRF models. 

\paragraph{Bottleneck Analysis.}
We first describe a general pipeline that unifies the computation flows across various NeRF models (\Sect{sec:bck}).
We then analyze the performance bottlenecks using this general pipeline.
This pipeline decomposes NeRF models into three key stages: Indexing, Feature Gathering, and Feature Computation. 
By analyzing the workloads across different NeRF models, we identify their inherent bottlenecks in both algorithm design and underlying hardware (\Sect{sec:motiv}).
Algorithmically, a NeRF model processes millions of rays, each containing hundreds of samples. Each of these samples requires executing a general matrix multiply (GEMM)-based inference, collectively imposing a significant computational load~\cite{mildenhall2021nerf}. 
Architecturally, the GEMMs associated with ray samples require fetching pre-trained features, leading to irregular memory accesses. This, in turn, results in irregular DRAM accesses and SRAM bank conflicts.

\paragraph{Algorithmic Support.}
\proj presents a plug-and-play extension to existing NeRF algorithms and significantly reduces the computational workload of NeRF models (\Sect{sec:algo}). 
Our method, \textit{sparse radiance warping} (\algo), reduces up to 88\% of radiance computations by reusing ray radiances in previously computed reference frames.
\algo operates based on the observation of \textit{radiance proximity}: the radiance of nearby rays originating from the same physical point is approximately equivalent. 
It is worth noting that \algo is not a replacement for current NeRF models; rather, it is a plug-and-play extension that can be integrated with all NeRF models.

\paragraph{Run-time Support.}
Chooinsg reference frames is critical to unleash the full power of \algo.
We observe that reference frames do not have to be a frame on the camera's past trajectory: reference frames merely provide pixel data used to warp target frames.
We introduce a proactive rendering run-time system (\Sect{sec:algo:dd}), which predicts optimal reference frames that are \textit{off} the camera trajectory and, thus, decouples the reference frame rendering from target frame rendering.
The run-time then schedules the reference and target frames to maximally overlap the rendering of the two sets of frames.


\paragraph{Dataflow Optimization.}
\algo reduces, rather than eliminates, computations in NeRF models. 
The remaining NeRF computations are still bounded by irregular DRAM accesses and frequent SRAM bank conflicts. 
Therefore, we propose two dataflow optimizations that enhance memory efficiency during radiance computation (\Sect{sec:mem}).
First, to mitigate irregular DRAM accesses, we propose to shift NeRF from a \textit{pixel-centric} order to a \textit{memory-centric} order. 
Our memory-centric order accesses the scene representations sequentially, inherently yielding full-streaming memory accesses.
Second, to tackle SRAM bank conflicts, we introduce a novel data layout strategy. Instead of the traditional feature-major data layout, which stores an entire feature vector within a single SRAM bank, we propose a channel-major layout where different channels of the same feature vector are distributed across multiple SRAM banks.

\paragraph{Hardware Augmentation.} To effectively support our two dataflow optimizations, we have enhanced the classic systolic array architecture with two significant augmentations. 
First, we extend the existing vector unit to support exponential computations, a necessity for the feature computation in 3DGS.
Secondly, we introduce a dedicated gathering unit designed specifically to accommodate our novel data layout. This new unit is tailored to manage the channel-major layout of data across multiple SRAM banks, ensuring that on-chip bank conflicts are eliminated.


The contributions of this paper are as follows. 
\begin{itemize}
    \item We introduce \algo, a novel algorithm that reduces up to 88\% of the GEMM in NeRF by exploiting radiance similarities between nearby rays.
    \item We propose a unified fully-streaming framework for NeRF that reduces the redundant DRAM access and eliminates SRAM bank conflicts, ensuring complete streaming DRAM accesses.
    \item We augment associated hardware supports on the existing accelerator to support our streaming algorithms, make it compatible across various NeRF models with structured and unstructured representations.
    \item We integrate \proj with six state-of-the-art NeRF solutions and demonstrate that \proj achieves a 53.1$\times$ and speed-up and 67.7$\times$ energy savings over a baseline with a dedicated DNN accelerator while maintaining less than 1.0~dB degradation in Peak Signal-to-Noise Ratio (PSNR).
\end{itemize}

\section{Background}
\label{sec:bck}

In this section, we start with reviewing the fundamentals in NeRF (\Sect{sec:back:basics}). Then, we propose a unified framework to express general NeRF algorithms (\Sect{sec:back:pipeline}). 

\subsection{NeRF Fundamentals}
\label{sec:back:basics}

The debate between neural rendering and traditional ray tracing (physically-based rendering)~\cite{pharr2023physically} is a widely contested topic in graphics. Our work does not seek to settle this debate; instead, we aim to enhance the efficiency of neural rendering, making it a more compelling option.

Fundamentally, NeRF rejuvenates the classic image-based rendering techniques~\cite{shum2008image} by learning the light-field~\cite{levoy2023light, gortler2023lumigraph} of a scene using modern deep learning methods.
NeRF does away with detailed 3D modeling and physically simulating light transport in space.
It avoids the complicated setup in ray tracing, e.g., modeling the geometry of the scene and describing material properties.
Instead, NeRF uses offline-captured images of a scene to train a differentiable model, which encodes the volume density and the light field in the scene (i.e., radiance of any ray)~\cite{levoy2023light, gortler2023lumigraph}.
At rendering time, given the camera pose where an image is to be rendered, the model is probed through the classic volume rendering~\cite{kaufman1993volume, levoy1988display} to synthesize the image.

Existing NeRF algorithms can be classified into two main categories: MLP-based models with structured representations~\cite{muller2022instant, sun2022direct, chen2022tensorf} and 3DGS with unstructured representations~\cite{kerbl20233d, lee2023compact, wu2024recent, chen2024survey}. Structured representations capture the geometry of the physical world using regular voxel grids, where the granularity of the voxels directly impacts the rendering quality and model size. 
On the other hand, unstructured representations are more flexible: the scene is represented by a point cloud and each point is, in turn, represented by an ellipse whose shape and color are learned in training.
The overall sizes of these unstructured representations are determined by two: the total number of Gaussian points and the parameter configurations of each Gaussian point.

\subsection{General NeRF Pipeline}
\label{sec:back:pipeline}

\begin{figure*}[t]
\centering
\includegraphics[width=\textwidth]{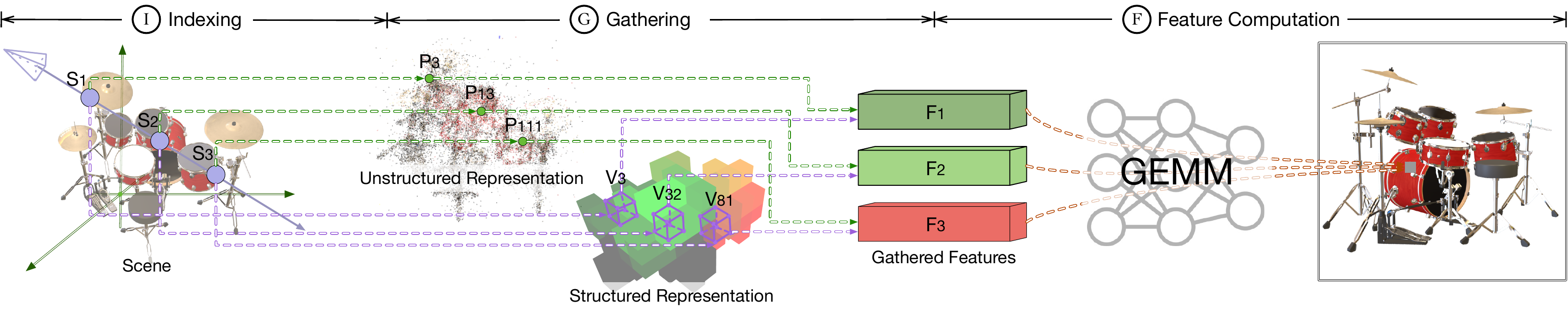}
\caption{The rendering pipeline of today's NeRFs consists of three stages: Indexing ($\cI$), Feature Gathering ($\cG$), and Feature Computation ($\cF$). For NeRFs with structured representations, each ray first samples points ($S_{1}$, $S_{2}$, and $S_{3}$) along the ray direction. Each ray sample gathers and interpolates 3D features from eight vertices of the intersected voxel ($V_3$, and $V_{32}$, and $V_{81}$) to obtain features ($F_1$, $F_2$, and $F_3$), as highlighted in purple. For NeRFs with unstructured representations, each ray directly intersects with Gaussian points ($P_3$, $P_{13}$, and $P_{111}$) to obtain features, as highlighted in green. Then, the features are fed into the GEMM-based computation to get the partial pixel values, as highlighted in orange. The final pixel value is summed from all partial pixel values~\cite{mildenhall2021nerf}.
}
\label{fig:algo_flow}
\end{figure*}

Despite numerous variants, we find that all prior NeRF algorithms can fit into a single computational paradigm. In this paradigm, the general NeRF pipeline consists of three stages: Indexing ($\cI$), Feature Gathering ($\cG$), and Feature Computation ($\cF$), as illustrated in \Fig{fig:algo_flow}.

\begin{figure}[t]
\centering
\begin{minipage}[t]{0.48\columnwidth}
  \centering
  \includegraphics[width=\columnwidth]{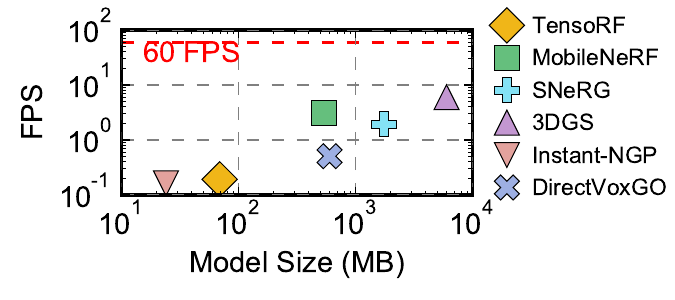}
  \caption{Frame rate vs. model size on the Xavier SoC~\cite{xaviersoc} across state-of-the-art NeRF algorithms\cite{muller2022instant, sun2022direct, chen2022tensorf, hedman2021baking, chen2023mobilenerf, kerbl20233d}.
  }
  \label{fig:compute_vs_model_size}
\end{minipage}
\hspace{2pt}
\begin{minipage}[t]{0.48\columnwidth}
  \centering
  \includegraphics[width=\columnwidth]{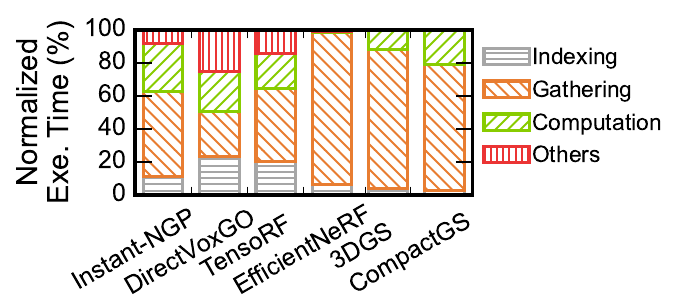}
  \caption{Normalized execution breakdown across state-of-the-art NeRF algorithms~\cite{muller2022instant, chen2022tensorf, hu2022efficientnerf, sun2022direct, kerbl20233d, lee2023compact}.}
  \label{fig:execution_dist}
\end{minipage}
\end{figure}

\paragraph{Indexing ($\cI$).}
NeRF renders images in a pixel-centric order. In this order, we first generate a ray for each pixel to be rendered.
Depending on the scene representation, rays have different fashions to sample points. With structured representation, each ray uniformly samples a set of points (e.g. $S_1$, $S_2$, and $S_3$ in \Fig{fig:algo_flow}) along its direction. 
In contrast, with an unstructured representation, each ray directly interacts with specific Gaussian points it intersects. 
For instance, in \Fig{fig:algo_flow}, ray would directly intersect with $P_3$, $P_{13}$, and $P_{111}$ at $S_1$, $S_2$, and $S_3$, respectively.
Any points that intersect with the ray contribute to the final pixel value.

\paragraph{Feature Gathering ($\cG$).} 
In this step, each ray sample gathers corresponding features for Feature Computation.
These features encode both the density and radiance field at the corresponding locations.
With structured representation, each ray sample would calculate the ID of the voxel that contains the sample.
Using the voxel ID, each ray sample finds the eight vertices of that voxel and gathers the features of the eight vertices.
In the example in \Fig{fig:algo_flow}, $S_1$ would access the eight vertex features in $V_3$.
This ray sample then computes its feature by trilinearly interpolating the feature vectors of the eight vertices.
With unstructured representation, each sample would directly apply the feature of the intersected Gaussian points.
For instance, $S_1$ would apply the feature at $P_{3}$.


\paragraph{Feature Computation ($\cF$).}
In Feature Computation, each ray sample passes its intermediate feature through a lightweight rendering model to obtain that sample's actual density and radiance value. 
Varied by designs, the major computation in the rendering model could be GEMM or other vector multiplications. For instance, Instant-NGP applies a lightweight MLP-based model to infer the radiance value while 3DGS leverages high-order spherical harmonics to approximate the radiance variation of a given point.
The final color of a ray (and thus the color of the pixel that is hit by the ray) would be computed by accumulating all ray samples along the ray direction.



\section{Motivation}
\label{sec:motiv}
In this section, we first profile the existing NeRF algorithms and identify the performance bottleneck, Feature Gathering, across NeRF algorithms (\Sect{sec:motiv:comp}). Next, we characterize the memory inefficiencies in Feature Gathering (\Sect{sec:motiv:mem}).

\subsection{Computation Inefficiencies}
\label{sec:motiv:comp}

\paragraph{Performance and Model Size.}
Today's NeRF models not only suffer from low performance but also present model storage challenges that could not fit in on-chip SRAMs.
\Fig{fig:compute_vs_model_size} shows the distribution of model size (x-axis) and frame rate (y-axis) for six state-of-the-art NeRF models~\cite{{muller2022instant, sun2022direct, chen2022tensorf, kerbl20233d, chen2023mobilenerf, hedman2021baking}} on a mobile Volta GPU~\cite{xaviersoc}. 
By overlaying the 60 FPS runtime requirement, current NeRF algorithms cannot achieve real-time rendering on mobile devices.

Moreover, the size of NeRF models greatly surpasses the SRAM capacities of a typical modern mobile SoC, which leads to frequent DRAM accesses. 
In particular, a significant bulk of the model size is constituted by the feature vectors of the voxels, which range from approximately 10 MB to 1,000 MB. In contrast, the model weights in Feature Computation are relatively small, usually between 10 KB and 100 KB. 

\paragraph{Performance Bottleneck.}
\Fig{fig:execution_dist} shows the execution breakdown of different stages across six popular NeRF algorithms on a mobile Volta GPU~\cite{xaviersoc}. All three stages take non-trivial execution time with Feature Gathering dominating the execution ($>$64\% of execution time on average).

\subsection{Memory Inefficiencies}
\label{sec:motiv:mem}

Given the memory-intensive nature of Feature Gathering, we characterize its memory access patterns. Although Feature Gathering benefits from computational parallelism (both across different rays and between samples on a single ray), it does not translate well to memory efficiency, leading to irregular DRAM accesses and frequent on-chip SRAM bank conflicts.

\begin{figure}[t]
\centering
\begin{minipage}[t]{0.48\columnwidth}
  \centering
  \includegraphics[width=0.8\columnwidth]{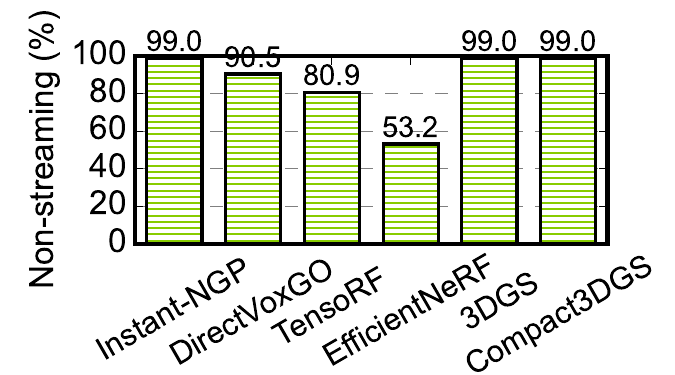}
  \caption{Percentage of non-continuous DRAM accesses in Feature Gathering $\cG$.}
  \label{fig:dram_non_stream}
\end{minipage}
\hspace{2pt}
\begin{minipage}[t]{0.48\columnwidth}
  \centering
  \includegraphics[width=0.8\columnwidth]{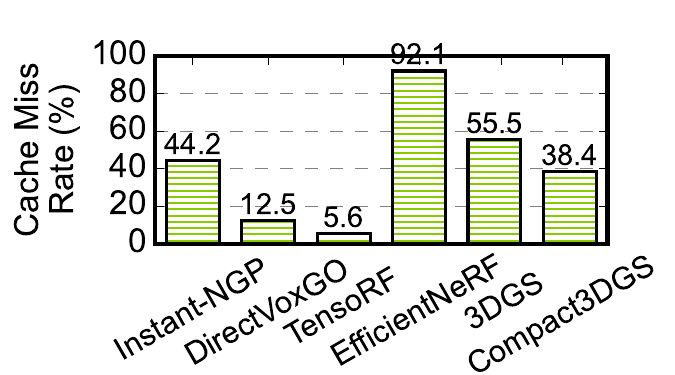}
  \caption{On-chip memory miss rate in Feature Gathering ($\cG$) across NeRF algorithms.}
  \label{fig:cache_miss}
\end{minipage}
\end{figure}

\paragraph{DRAM Access Inefficiency.} The inefficiency in DRAM access can be attributed to two main factors: non-streaming DRAM access and redundant DRAM access, both stemming from the inherent \textit{pixel-centric} rendering in NeRF models.

NeRF inference is parallelized across pixels as shown in \Sect{sec:back:pipeline}. 
This pixel-centric rendering approach, however, introduces two distinct levels of memory-access irregularities: inter-ray and intra-ray irregularity.
Inter-ray irregularity is from that two rays corresponding to adjacent pixels might access non-continuous memory regions. Despite originating from spatially close points, these rays may diverge as they travel through space, leading to varied memory access patterns.

Intra-ray irregularity arises when sampling points along a single camera ray access discontinuous memory regions. 
This occurs because the feature vectors corresponding to different ray samples may be stored at arbitrary locations in memory.
As shown in \Fig{fig:algo_flow}, for unstructured representations, ray samples, $S_1$ and $S_2$, intersect points, $P_2$ and $P_{13}$; for structured representations, ray samples, $S_1$ and $S_2$, intersect voxels, $V_3$ and $V_{32}$.
In both scenarios, gathered features are spatially separate and stored discontinuously in DRAM.
\Fig{fig:dram_non_stream} shows the non-streaming DRAM access in six popular NeRF algorithms.
Over 87\% of DRAM access is non-streaming.
Particularly, Instant-NGP, 3DGS, and Compact3DGS all have high non-streaming DRAM access ($>$99\%) due to irregular memory access to their naive feature representations.

Given that the entire feature representations cannot be stored completely on-chip (\Fig{fig:compute_vs_model_size}),
both types of irregularities would lead to repetitive redundant accesses to each voxel during rendering, which translates to redundant DRAM accesses.
Given an on-chip buffer size of 2 MB with oracle replacement~\cite{xaviersoc}, 
\Fig{fig:cache_miss} presents the cache miss rates for various NeRF algorithms during Feature Gathering. 
The miss rate can be as high as 92\% (average 41\%).
Practically, an even smaller on-chip buffer is expected to accommodate feature representations, further degrading rendering performance.

\begin{figure}[t]
\centering
\begin{minipage}[t]{0.35\columnwidth}
  \centering
  \includegraphics[width=\columnwidth]{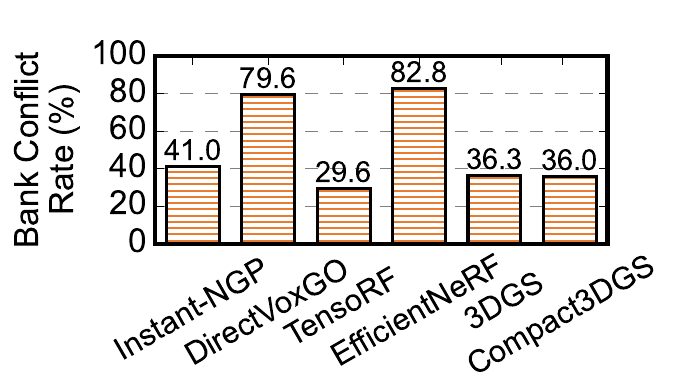}
  \caption{SRAM bank conflict rate in Feature Gathering ($\cG$), assuming 16 banks and 16 concurrent ray queries.}
  \label{fig:sram_conflict}
\end{minipage}
\hspace{2pt}
\begin{minipage}[t]{0.35\columnwidth}
  \centering
  \includegraphics[width=\columnwidth]{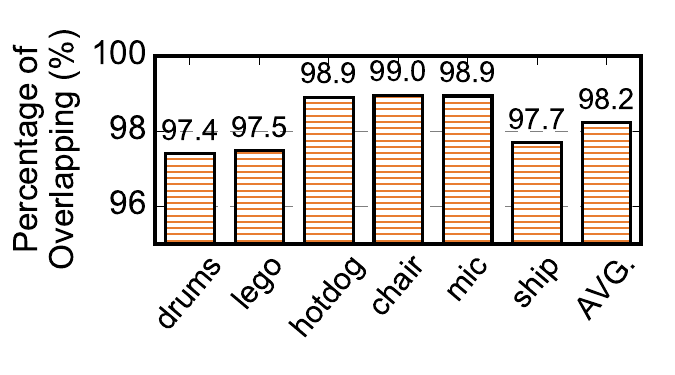}
  \caption{The overlapping percentage across six scenes in Synthetic-NeRF~\cite{mildenhall2021nerf}.}
  \label{fig:warp_pct}
\end{minipage}
\hspace{2pt}
\begin{minipage}[t]{0.25\columnwidth}
  \centering
  \includegraphics[width=\columnwidth]{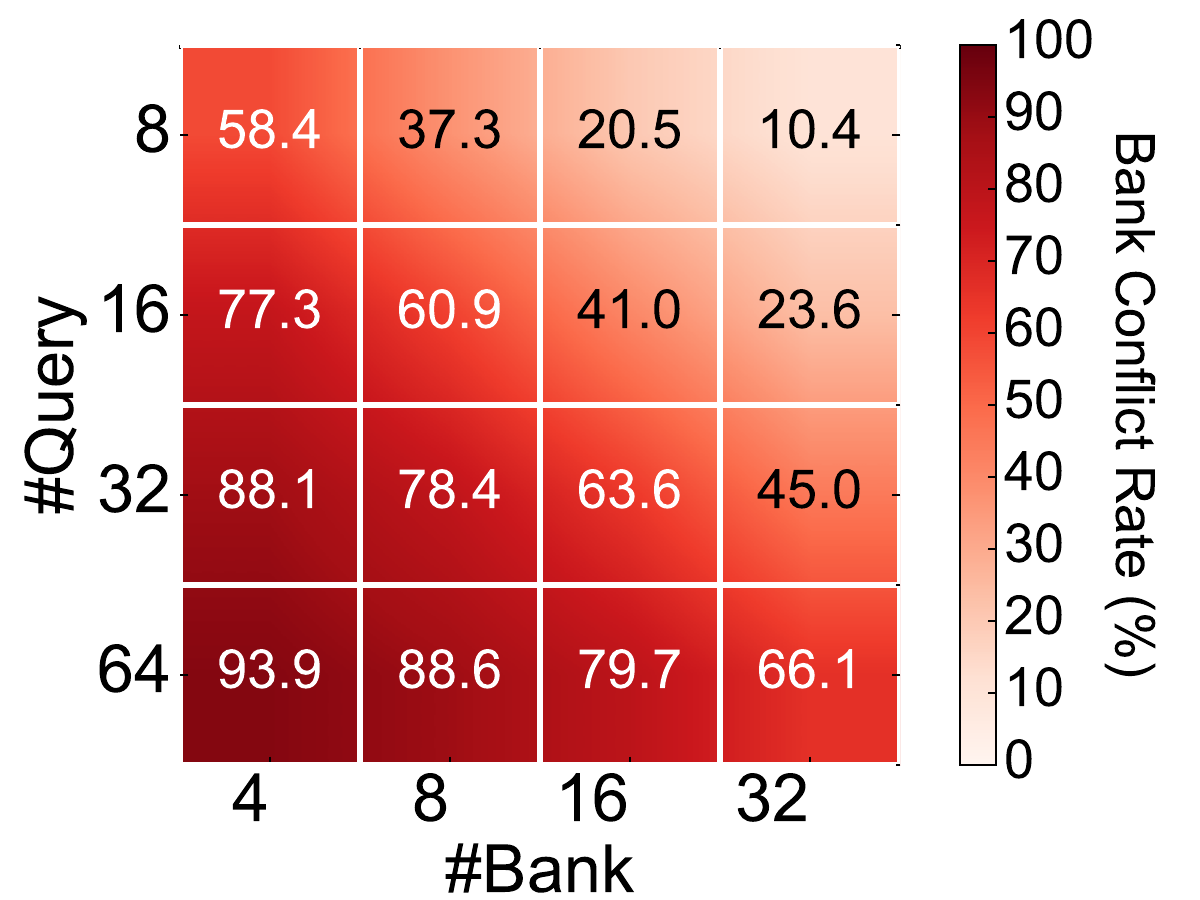}
  \caption{Profiling SRAM bank conflicts of Instant-NGP with various banks and concurrent queries.}
  \label{fig:instance_ngp_profile}
\end{minipage}

\end{figure}

\paragraph{SRAM Access Inefficiency.}
On-chip memory access in NeRF results in frequent bank conflicts.
In conventional DNNs, the memory access patterns can be determined statically. Thus, bank conflicts can be eliminated through meticulous data layout across SRAM banks~\cite{zhou2021characterizing, kirk2016programming}.
Conversely, the data access pattern of Feature Gathering in NeRF depends on the camera view and cannot be known offline. 

\Sect{sec:mem:bank} will provide a detailed description of the causes of bank conflicts in Feature Gathering. Here, we simply show the bank conflict rate of Feature Gathering across NeRF algorithms (\Fig{fig:sram_conflict}). Assuming a 2 MB buffer with 16 banks and 16 concurrent camera rays, the average bank conflict rate is 51\%, with EfficientNeRF reaching as high as 83\%.
A larger number of concurrent rays would lead to a higher bank conflict rate. For instance, the bank conflict rate of Instant-NGP increases to 80\% when the number of rays escalates to 64 (\Fig{fig:instance_ngp_profile}).
Increasing the number of banks does alleviate bank conflicts. However, heavily banked SRAM designs are highly undesirable due to costly crossbars~\cite{agarwal2009garnet, grot2011kilo}.




\section{Sparse Radiance Warping}
\label{sec:algo}

\no{This section introduces \textit{sparse radiance warping} (\algo), an algorithm that exploits the radiance similarity across rays from nearby camera views.
We first provide an intuition of \algo algorithm (\Sect{sec:algo:intuition}), followed by a description of the overall algorithm (\Sect{sec:algo:main}).
Lastly, we discuss a customized runtime to support \algo (\Sect{sec:algo:dd}).
}


\begin{figure}[t]
  \centering
  \includegraphics[width=0.85\columnwidth]{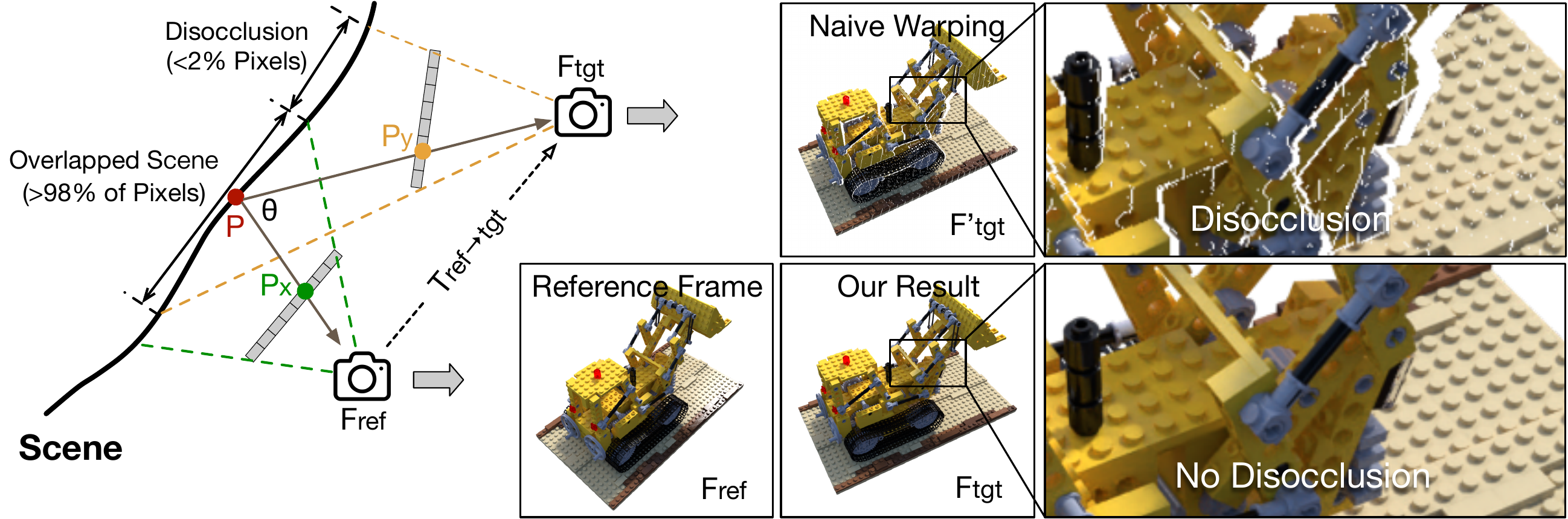}
  \caption{Intuition of our \algo algorithm. The radiance of the $\overline{P P_x}$ ray can approximate the radiance of the $\overline{P P_y}$ ray if the angle $\theta$ between these two rays is sufficiently close. Leveraging this intuition, we show examples of a reference frame $F_{ref}$, a result of naive warping $F'_{tgt}$ and our result $F_{tgt}$ by \algo. Note that disocclusions (missing pixels) are eliminated in our result.}
  \label{fig:intuition}
\end{figure}

\subsection{Intuition}
\label{sec:algo:intuition}

The goal of \algo is to reuse pixel values rendered in previous frames using a technique called image warping.
\Fig{fig:intuition} illustrates our idea, which starts from a previously rendered frame, called a reference frame, $F_{ref}$.
For a given pixel in $F_{ref}$, say $P_x$, we can find the point in the scene $P$ that is captured by that pixel.
When rendering a new target frame $F_{tgt}$, the same point $P$ is captured as a new pixel $P_y$ in the target frame $F_{tgt}$.
The assumption here is that if the camera poses of $F_{tgt}$ and $F_{ref}$ are sufficiently close, 
the radiance of both $\overline{P P_x}$ ray and $\overline{P P_y}$ ray are approximately similar.
Thus, the pixel value $P_x$ can be simply reused in $P_y$, avoiding rendering $P_y$ through the compute-intensive NeRF model.


This warping idea avoids rendering pixels in $F_{tgt}$ whose corresponding scene points are also captured by $F_{ref}$.
The larger the overlap between $F_{ref}$ and $F_{tgt}$ is, the less NeRF computation is required.
\Fig{fig:warp_pct} characterizes the overlapping between two adjacent frames in the Synthetic-NeRF dataset~\cite{mildenhall2021nerf}.
More than 98\% of pixels are overlapped (standard deviation: 1.7\%), indicating that less than 2\% of pixels require re-rendering.
The same conclusion holds for real-world datasets:
on the Unbounded-360~\cite{barron2022mipnerf360} and Tanks and Temples~\cite{Knapitsch2017} datasets, only 4.3\% and 4.9\% pixels cannot be warped, respectively.
The high overlap is not an artifact of a particular dataset but a fundamental attribute of real-time rendering, where consecutive frames are necessarily in close proximity because the observer/camera does not jump arbitrarily.

The non-overlapped pixels, called \textit{disoccluded pixels}, arise when the previously occluded scene in $F_{ref}$ becomes visible in $F_{tgt}$.
\Fig{fig:intuition} shows the effect of a naive warping.
Without recalculating the disoccluded pixels, the rendered image $F'_{tgt}$ has clear ``holes'', because the disoccluded pixels cannot be warped from the reference frame.
Our idea then is to calculate the disoccluded pixels using the original NeRF model, which now renders only a small amount of (e.g., 2\%) sparsely disoccluded pixels in the target frame.


\subsection{Basic Algorithm}
\label{sec:algo:main}

\begin{figure*}[t]
\centering
\includegraphics[width=\columnwidth]{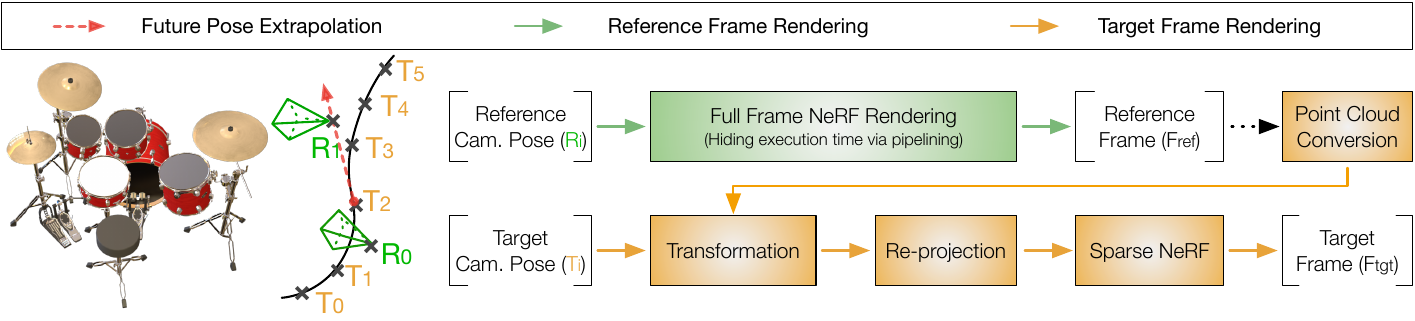}
\caption{An overview of the \algo algorithm.
Only reference frames ($R_i$) undergo full-frame NeRF inference as denoted by the green path.
All target frames ($T_i$) are computed using the lightweight warping operations denoted by the orange path.
The reference frames are not on the camera trajectory so reference frame rendering and target frame rendering can be overlapped.
Camera poses at reference frames are extrapolated using the previous camera poses.
}
\label{fig:algo}
\end{figure*}

In \algo there are two rendering paths, which are illustrated in \Fig{fig:algo}: a compute-intensive path (in green) to render reference frames ($R_0$ and $R_1$) using full-frame NeRF rendering, and a lightweight path (in orange) that uses the warping idea in \Sect{sec:algo:intuition} to render target frames ($T_0$ -- $T_5$).
We first describe how to warp from a reference frame to a target frame, then discuss the choice of reference frames.

\paragraph{Target Frame Rendering.} 
There are four steps in rendering a target frame: \circled{white}{1} point cloud conversion, \circled{white}{2} transformation, \circled{white}{3} re-projection and \circled{white}{4} sparse NeRF rendering. 

\circled{white}{1} Given a reference frame $F_{ref}$, we first convert $F_{ref}$ into a point cloud $P_{ref}$, which represents the 3D scene in the reference camera coordinate system.
The transformation uses scene depth and the camera's intrinsic parameters. Mathematically, it can be expressed as follows:
\begin{gather}
    P_{ref} = \begin{bmatrix}
        \frac{D_{ref}}{f} & 0 & -\frac{D_{ref} C_x}{f} \\
        0 & \frac{D_{ref}}{f} & -\frac{D_{ref} C_y}{f} \\
        0 & 0 & D_{ref}
    \end{bmatrix} \times F_{ref}
\end{gather}
where $D_{ref}$ is the depths of points in $P_{ref}$ corresponding to pixels in $F_{ref}$; $D_{ref}$ can be obtained through a standard rasterization pipeline (using depth buffer)~\cite{shirley2009fundamentals};
$f$ is the camera focal length;
[$C_x$, $C_y$] is the camera center.
Both focal length and camera center are part of the camera's intrinsic parameters~\cite{forsyth2002computer}.

\circled{white}{2} The $P_{ref}$ calculated so far is expressed in the coordinate system of the reference frame.
To render the target frame, we must transform the point cloud to the coordinate system of the target frame --- using a simple linear transformation:
\begin{equation}
    P_{tgt} = T_{ref \rightarrow tgt} \times P_{ref}
\end{equation}
where $T_{ref \rightarrow tgt}$ is the transformation matrix between the reference camera pose $R_i$ and the target camera pose $T_i$;
$P_{tgt}$ denotes the point cloud in the target frame's coordinate system.

\circled{white}{3} Once we have $P_{tgt}$ expressed at the current camera coordinate system, obtaining the frame at the current camera pose requires a standard perspective projection in the classic rasterization pipeline~\cite{shirley2009fundamentals}:
\begin{gather}
    F'_{tgt} = \begin{bmatrix}
        \frac{f}{D_{tgt}} & 0 & 0 & C_x\\
        0 & \frac{f}{D_{tgt}} & 0 & C_y \\
        0 & 0 & \frac{1}{D_{tgt}} & 0
    \end{bmatrix}  \times P_{tgt}
\end{gather}
\noindent where $D_{tgt}$ is the depth of all the points corresponding to the pixels in the target frame.
A standard depth buffer technique (also known as a z-buffer) is used to maintain the correct depth of objects in 3D space from the camera perspective.

\circled{white}{4}
As shown in \Fig{fig:intuition}, naively warped frame $F_{tgt}'$ contains disocclusion artifacts.
To mitigate disocclusions, we simply run the original NeRF model for those disoccluded pixels,
\begin{equation}
    F_{tgt} = F'_{tgt} \circledast \Gamma_{sp}
\end{equation}
\noindent $\Gamma_{sp}$ denotes sparse NeRF rendering of disoccluded pixels, and $\circledast$ combines the warped pixels with NeRF-rendered pixels.

Interestingly, ``holes'' in a target frame can be attributed to two factors: disocclusion and void (i.e., areas in the scene where there is nothing).
To avoid unnecessary computation on the latter, we perform a simple depth test so that pixels whose depth is infinite are skipped in sparse NeRF rendering.
The depth map of the current frame can be obtained through, again, the standard perspective projection.
The overhead of such a projection is minimal.
In our measurement, the latency of processing, e.g., one million points, is less than one millisecond on a Nvidia Volta mobile GPU.

\subsection{Proactive Rendering Runtime}
\label{sec:algo:dd}

\begin{figure}[t]
\centering
\includegraphics[width=\columnwidth]{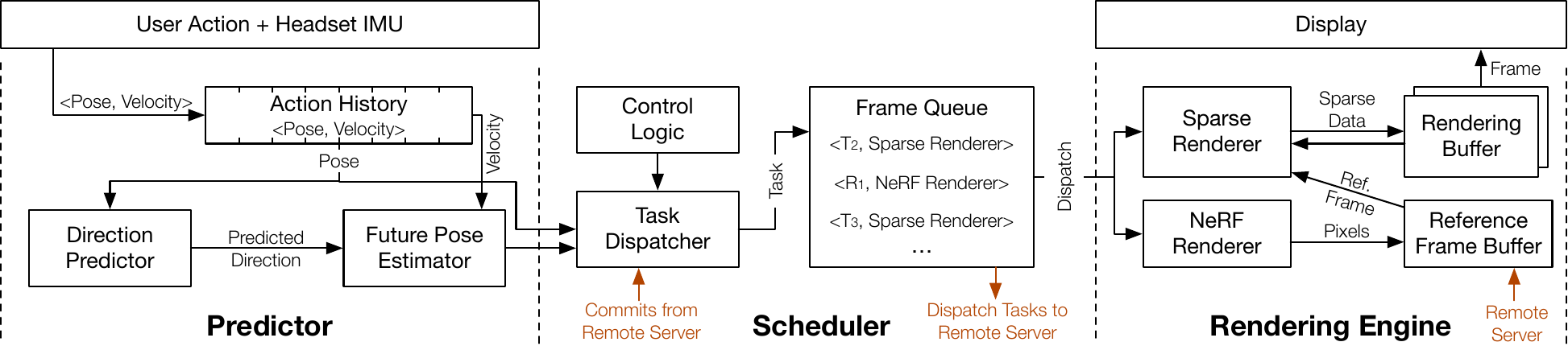}
\vspace{-5pt}
\caption{Proactive Rendering Runtime consists of three key components: a predictor, a scheduler, and a rendering engine. The predictor proactively predicts the future camera poses for reference frames, and the predicted poses ($R_i$) are sent to the task dispatcher along with real-time camera poses ($T_i$) for target frames. The scheduler then dispatch tasks to the corresponding renderers to achieve concurrent rendering for reference and target frames in \Fig{fig:algo}. Our run-time also allows Reference frames to be offloaded to a remote server by leveraging cloud computing resources. Once the reference frames are finished, the remote server will send them back to Reference Frame Buffer.
}
\label{fig:schedule}
\end{figure}

In \algo, the rendering of a target frame depends on the existence of a reference frame. The choice of reference frames and when to render them affect both the rendering quality and the performance.
Rendering reference frames on the trajectory is a common strategy in previous work~\cite{zhu2018euphrates, feng2019asv, song2020vr, buckler2018eva2}.
The idea is that adjacent frames are highly correlated so one can extrapolate from a previously rendered frame to render the current frame.
While this approach is work-efficient since reference frames are required to be rendered regardless, it has an inherent limitation in which it poses the data dependency between reference frames and target frames. Specifically, a reference frame can only be rendered after the completion of the previous target frame. This dependency inherently forces the rendering process into a serial order, where target frames and reference frames cannot be rendered concurrently, potentially slowing down overall system performance.

Our key observation is that a reference frame can be any frame, even those that are not on the camera's trajectory, such as $R_0$ and $R_1$ in \Fig{fig:algo}.
As long as the reference frame is close to the camera trajectory, the radiance approximation still holds.
We introduce a run-time system that predicts suitable reference frames \textit{off the trajectory}; that way, we can overlap reference frame rendering with target frame rendering.
As depicted in \Fig{fig:schedule}, our runtime system comprises three key components: a predictor, a scheduler, and a rendering engine.

\paragraph{Predictor.}
We introduce a \textit{proactive} approach to predict camera poses for \textit{future} reference frames.
The idea is that reference frames might not be the frames viewed by users. Rather, the primary functionality of reference frames is to supply pixel information for target frames. To this end, we propose a lightweight direction predictor that utilizes a brief history of past camera poses to predict the future direction of the current trajectory.
In particular, our direction predictor takes the five previous camera poses as input and uses a Bayesian Ridge regression model~\cite{tipping2001sparse} to estimate the upcoming camera direction. Experimental results show that our predictor maintains a low prediction error ( $<$0.01 rad) with a negligible runtime overhead.

Once the camera direction is predicted, we use the position of the last two rendered frames, $T_1$ and $T_2$, to calculate the current velocity $v$ at $T_2$,
$v = \frac{T_2 -  T_1}{\Delta t}$, where $\Delta t$ represents the interval between two consecutive frames.
We then calculate the pose of the reference frame, $R_1 = T_2 + v \times t_r,~~~t_r = \frac{N}{2}\Delta t$, where $N$ is the number of target frames that share the same reference frame (i.e., 4 in this example as $T_2$ -- $T_5$ share $R_1$). $\Delta t$ represents the interval between two consecutive frames.
Using $\frac{N}{2}$ allows the reference frame to be roughly at the center of, and thus increase the overlap with, its target frames.

\paragraph{Scheduler.} 
Both predicted and current camera poses are sent to the task dispatcher, which inserts rendering tasks into the frame queue. 
Depending on the type of camera pose, the rendering tasks will be dispatched to the corresponding renderers. 
For instance, a predicted camera pose ($R_1$) would invoke a full-frame rendering task to NeRF Renderer while a target camera pose $T_3$ will be dispatched to Sparse Renderer \textit{simultaneously}. 
Our strategy can effectively hide the latency of reference frame rendering by overlapping with rendering multiple target frames. 
Our experiment shows that a single reference frame can effectively support up to 30 target frames with minimal loss in visual quality (\Sect{sec:eval:disc}).

Meanwhile, a control logic continuously monitors the task scheduling to guarantee two properties.
First, the control logic would check the data dependencies among the rendering tasks so that no target frame rendering would be dispatched without its corresponding reference frame rendered.
Second, if the angle ($\theta$) subtended by a ray in the reference frame and its corresponding ray in the target frame exceeds a predefined threshold ($\phi$), the control logic would stop the frame warping process and initiate a full-frame NeRF rendering task instead. 
This ensures that rendering quality is guaranteed with no significant delays. 

\paragraph{Rendering Engine.} 
Once a rendering task is dispatched, the rendering engine activates a corresponding renderer based on the task type.
For example, a reference frame rendering task is directed to the NeRF renderer, which activates full-frame NeRF rendering to generate a reference frame and stores it in the reference frame buffer. Here, we maintain a dedicated reference buffer, enabling the sparse renderer and the NeRF renderer to execute concurrently without mutual interference.
Meanwhile, leveraging the previously generated reference frame, the sparse renderer performs sparse NeRF rendering along with other lightweight computations shown in \Fig{fig:algo}.
To facilitate efficient processing, the rendering buffer is double-buffered, allowing it to hold some intermediate data during the execution of \algo.

\paragraph{Support Remote Rendering.}
Our runtime system also supports offloading the heavy full-frame NeRF rendering to a remote server and receives the rendered result from the remote server as highlighted in \Fig{fig:schedule}. 
The end of \Sect{sec:exp} describes the rationale of remote rendering, and \Sect{sec:eval:perf} quantifies the advantage of \algo in remote rendering.

\section{Memory Optimizations}
\label{sec:mem}

While \algo reduces NeRF computation of target frames, reference frames still execute full-frame NeRF rendering, which is bottlenecked by redundant and irregular DRAM accesses of feature vectors and the frequent on-chip bank conflicts as shown in \Sect{sec:motiv:mem}.
This section first describes an algorithmic optimization that eliminates redundant DRAM accesses and guarantees fully-streaming DRAM accesses (\Sect{sec:mem:fs}).
We then discuss a new on-chip data layout that eliminates SRAM bank conflicts (\Sect{sec:mem:bank}).

\subsection{Fully-Streaming NeRF Rendering}
\label{sec:mem:fs}

\begin{figure*}[t]
\centering
\includegraphics[width=\columnwidth]{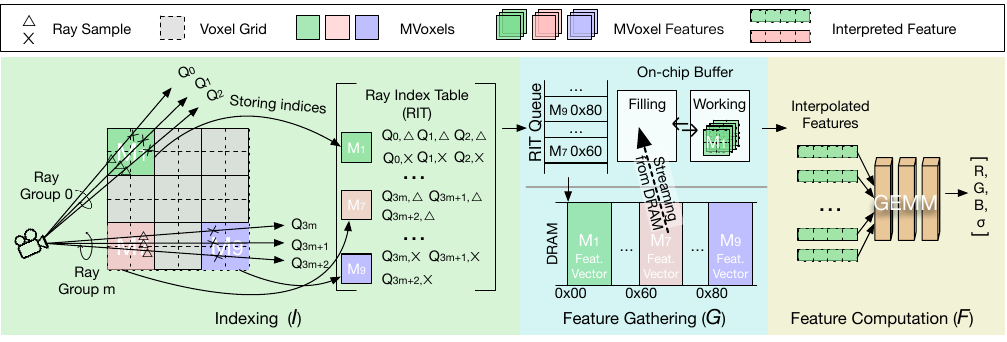}
\vspace{-5pt}
\caption{Fully-streaming NeRF rendering algorithm for structured representations~\cite{muller2022instant, sun2022direct, chen2022tensorf}. We first group all the voxels into MVoxels, which are continuously stored in the DRAM.
The Ray Index Table (RIT) records, for each MVoxel, the IDs of ray samples whose features reside in the MVoxel.
During feature gathering ($\cG$), the entries in the RIT are sequentially accessed, essentially streaming the MVoxels from the DRAM.
Each time an MVoxel is loaded on-chip, we process all the ray samples whose feature vectors are in that MVoxel.
The Feature Computation stage is unchanged.
}
\label{fig:streaming}
\end{figure*}




\paragraph{Architectural Assumptions.}
We assume a DNN accelerator for MLP operations.
The accelerator has an on-chip buffer (scratchpad) to store 3D feature vectors (either voxel features in structured representations or point features in unstructured representations)  in the NeRF model.
However, this buffer is generally too small (1~MB -- 3~MB) to hold all feature vectors (10~MB -- 1000~MB).
Additionally, there is a dedicated on-chip buffer to store NeRF model weights; these weights are generally small (10~KB -- 100~KB).



\paragraph{Memory-Centric Rendering.}
We propose a \textit{memory-centric} rendering, where the order of computation is based on where the ray samples reside in the DRAM.
At a high level, we sequentially read, in chunks, feature vectors that are contiguously laid in memory.
As each chunk is loaded to the on-chip SRAM, we render the ray samples whose feature vectors happen to reside in the chunk.
We throw away the chunk only after all the associated ray samples have been computed.
In this way, we guarantee that each feature vector is read only once and the DRAM accesses to the feature vectors are fully streaming.
Memory-centric rendering incurs no additional storage overhead. The vertex features are stored in memory \textit{as is}, without duplication.
What is being reordered is the feature \textit{accessing} order.
\Fig{fig:streaming} illustrates this idea and how the three stages in NeRF are changed accordingly.

\paragraph{Indexing ($\cI$).}
We first group spatially close feature vectors into ``macro voxels'' (MVoxels).
For instance, with structured representation, in \Fig{fig:streaming}, we can combine every $2\times2$ voxels into one MVoxel; with unstructured representation, we group all point features confined within one MVoxel.
All the data in a MVoxel is loaded to the SRAM together when the MVoxel is loaded.
The intention here is that, via offline analysis, we guarantee that the overall data stored in one MVoxel is smaller than the on-chip buffer size.
To better exploit memory locality, features within one MVoxel are stored continuously in the DRAM, the same as MVoxels.

We then compute a Ray Index Table (RIT), where each MVoxel has an entry.
Each entry records the IDs of all the ray samples whose features reside in that particular MVoxel.
Note that the ray sample-to-voxel mapping has to be calculated in the original NeRF models too; we simply group all such calculations and store the results in a table.


\paragraph{Feature Gathering ($\cG$) and Feature Computation ($\cF$).} 
During gathering, we load the MVoxels from the DRAM to the SRAM sequentially.
When an MVoxel is loaded, we look up the RIT to find all the ray samples that can be computed.
A standard double buffer is used here to overlap MVoxel loading and the on-chip computation.
Feature Computation remains unchanged compared to the baseline NeRF rendering.

In practice, we first group a set of nearby rays into a ray group. For each ray group, we identify MVoxel IDs that all ray samples in the ray group would index to, from the near-camera point to the far distance. Based on the indexed MVoxel IDs, these ray samples are gathered into a ray index table (RIT). The Feature Computation is also performed starting from the near-camera MVoxel, thus, the early ray termination in the general NeRF rendering still applies.





\paragraph{Accommodating Unstructured Representation.} The naive approach to adapting an unstructured representation is, at Indexing ($\cI$), to uniformly partition the entire scene space into the same-size MVoxels while ensuring that the point features within one MVoxel can fit in the on-chip buffer. However, this uniform partitioning often leads to significant under-utilization of the on-chip buffer. The root cause is that Gaussian points are not evenly distributed across the space, so there would be many MVoxels that contain no Gaussian points.

\begin{figure}[t]
\begin{minipage}[t]{0.71\columnwidth}
    \centering
  \includegraphics[width=\columnwidth]{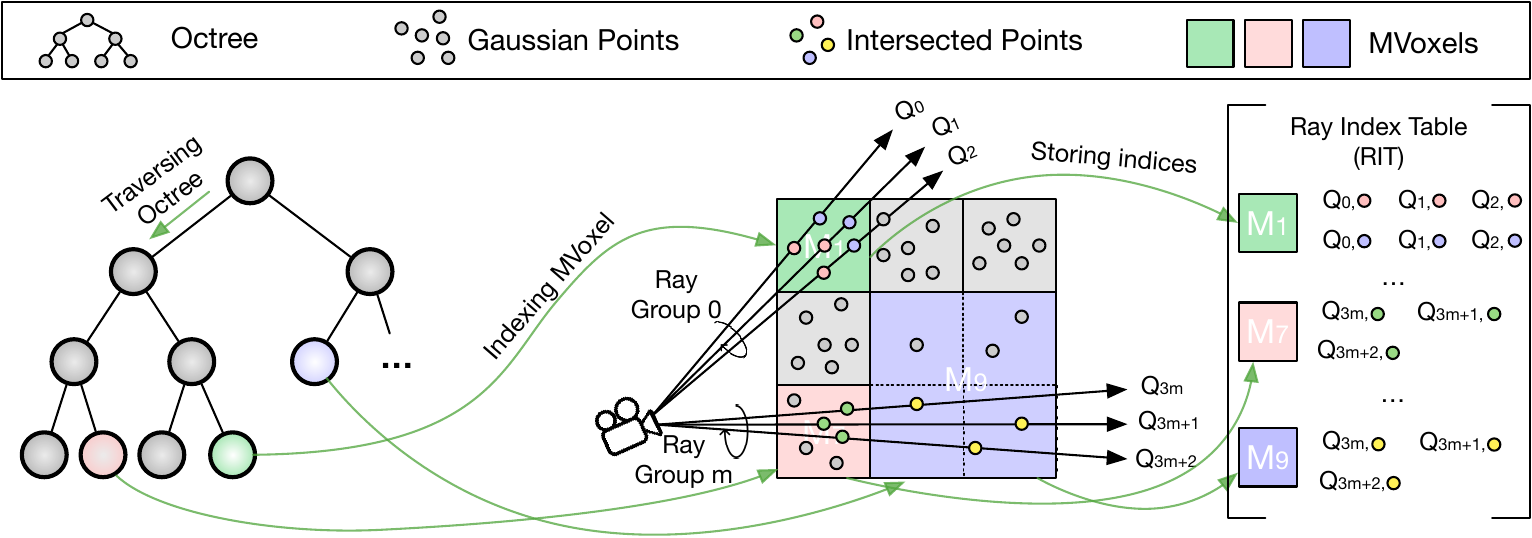}
  \caption{Modified Indexing ($\cI$) stage for unstructured representations~\cite{kerbl20233d, wu2024recent, chen2024survey}. We merge adjacent low-density MVoxels for better on-chip buffer utilization. An offline-built octree is used to help MVoxel traversal.}
  \label{fig:point_indexing}
\end{minipage}
\hspace{2pt}
\begin{minipage}[t]{0.26\columnwidth}
  \centering
  \includegraphics[width=\columnwidth]{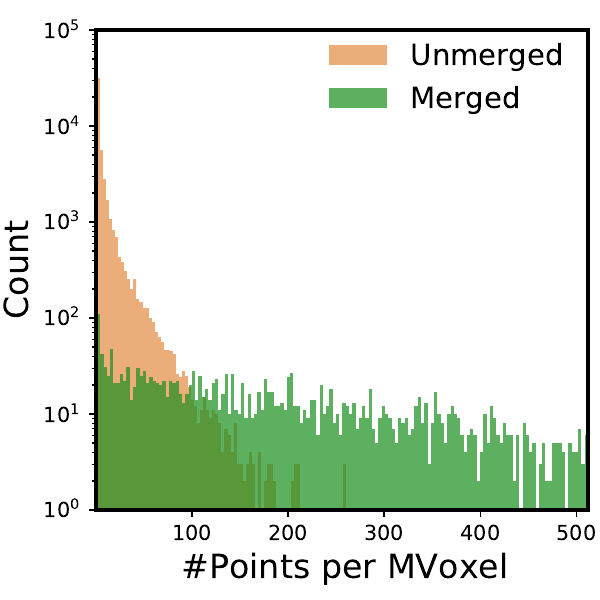}
  \caption{Distribution of points per MVoxel before and after merging. Post-merging octree improves on-chip buffer utilization.}
  \label{fig:octree_dist}
\end{minipage}
\end{figure}

To improve the utilization of the on-chip buffer, we merge adjacent MVoxels with low point densities and process them together. \Fig{fig:point_indexing} illustrate our idea. For instance, given that the four purple MVoxels (at the bottom right) are sparser than their neighboring MVoxels, we group the them into one single, larger MVoxel $M_9$. In our implementation, we balance the complexity and efficiency and only groups of $2\times2\times2$ MVoxels into a larger MVoxel if the combined size of the point features within these eight MVoxels does not exceed the capacity of the on-chip buffer. The remaining two operations, Feature Gathering and Feature Computation, are kept the same. 

Coupled with variable-sized MVoxels, we use an octree structure to facilitate traversing and processing these MVoxels.
As \Fig{fig:point_indexing} illustrates, starting from the root node, the tree traversal would first visit one of its subtree nodes and perform a depth-first-search until a leaf node is reached. A lead node represents a MVoxel. The possible rays would perform Gaussian point-ray intersections to check which Gaussian points would intersect with the rays and store the results in RIT. This process would iterate until all MVoxels are visited. 

Note that we group low-density MVoxels into larger MVoxels (eight subtree nodes into one single node), resulting in an octree that is not perfectly balanced. However, this unbalanced octree leads to more efficient use of the on-chip buffer. 
\Fig{fig:octree_dist} shows the point-per-MVoxel distribution before and after merging. Post-merging, the average number of points per MVoxel improves significantly from 7.2 to 172.8, indicating a more compact and memory-efficient structure. \Sect{sec:eval:perf} will further illustrate the performance benefits of the merged octree.

\paragraph{Accommodating Hierarchical Data Encodings.}
Some NeRF algorithms, instead of storing voxel features directly, use hierarchical data structures such as hierarchical voxel grid~\cite{muller2022instant}, hashing~\cite{olszewski2023hashcc}, and factorized tensor~\cite{chen2022tensorf} to encode.

To accommodate our fully-streaming data flow with these hierarchical data structures, we first partition 3D voxels at each level into MVoxel grids.
For hash grid data structures, each hash entry is a MVoxel rather than one single voxel feature so that DRAM streaming within one MVoxel can be guaranteed.
During Feature Gathering, we group all rays into small ray groups (\Fig{fig:streaming}) and collect features level-by-level for a given ray group.
Once we have traversed all levels, we can then compile all the vertex features necessary for this ray group.
When the 3D voxel dimensions in the last several levels are too large, loading each MVoxel entirely would lead to low utilization of voxels, wasting DRAM bandwidth.
In that case, we revert back to the original (non-streaming) data flow.
This reversion happens in, for instance, Instant-NGP~\cite{muller2022instant} from level 5 (out of 8 levels) onwards.
As a result, about half of the DRAM traffic on Instant-NGP is non-streaming (which is faithfully captured in evaluation).

\begin{figure}[t]
\centering
\subfloat[The feature vector layout in existing NeRF accelerators, where vertex features are spread across SRAM banks, but all the channels in the same feature vector are stored in the same bank.]{
    \label{fig:bank_conflict_exp}	
    \includegraphics[width=0.48\columnwidth]{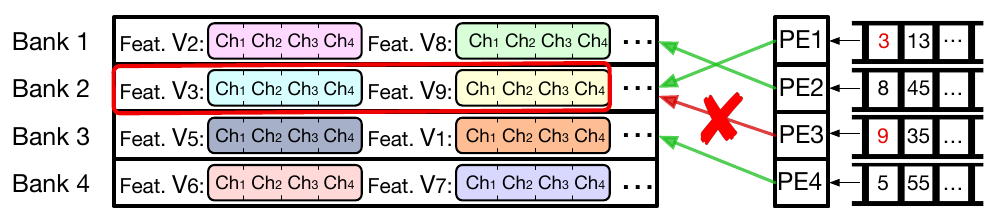} }
\hspace{5pt}
\subfloat[Our data layout spreads channels of a feature vector across different banks.]{
    \label{fig:our_layout}
    \includegraphics[width=0.48\columnwidth]{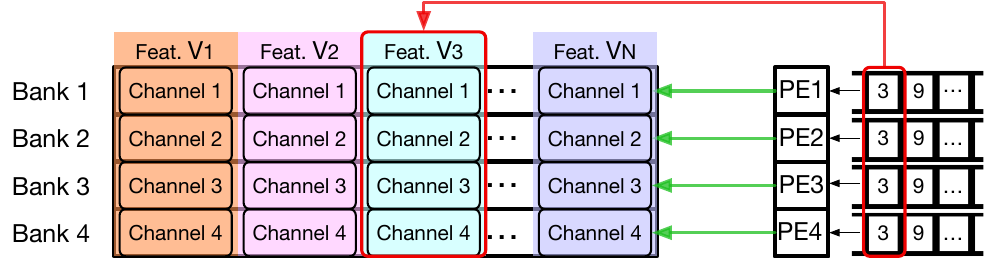} } 
\caption{A comparison between the original feature vector layout (\Fig{fig:bank_conflict_exp}) and our data layout (\Fig{fig:our_layout}).
The example has four banks and four concurrent PEs.
In \Fig{fig:bank_conflict_exp}, a bank conflict occurs when PE$_1$ and PE$_3$ (each collecting features for a different ray sample) access two different features from bank 2.
Our data layout (\Fig{fig:our_layout}) eliminates bank conflicts by 1) spreading channels across banks and 2) having each PE collect a particular channel across different ray samples.}
\label{fig:bank_conflict_comp}
\end{figure}

\subsection{Bank Conflict-Free Interleaving}
\label{sec:mem:bank}

With fully-streaming DRAM accesses, the inefficiency shifts to the on-chip SRAM, which experiences frequent bank conflicts (\Fig{fig:sram_conflict}),
which arise when different rays access the vertex features located at the same bank, leading to stalls in Feature Gathering ($\cG$). 
Critically, unlike conventional DNNs where one can orchestrate data layout offline to avoid bank conflicts~\cite{zhou2021characterizing, kirk2016programming}, the SRAM access pattern of ray samples is known only at the run time, because the exact ray samples depend on the run-time camera pose information.


The reason behind bank conflicts in Feature Gathering has to do with the way feature vectors are laid out in SRAM; \Fig{fig:bank_conflict_exp} illustrates this point.
State-of-the-art NeRF accelerators~\cite{li2023instant, lee2023neurex} store feature vectors in the SRAM using a \textit{feature-major} order, where all the channels of a feature vector are stored in the same SRAM bank.
Assume in this example we have four PEs, each responsible for collecting the feature vector for a particular ray sample.
PE$_1$ and PE$_3$ are responsible for two ray samples, which require feature vectors 3 and 9, respectively.
However, the two feature vectors happen to reside in the same bank, causing a bank conflict.

To address this issue, we propose a \textit{channel-major} layout, as illustrated in \Fig{fig:our_layout}, where different channels of the same feature vector are spread across banks.
For instance, bank 1 stores the first channel of all feature vectors within one MVoxel.
In cases where the feature channel size exceeds the bank size, the storing sequence restarts from bank 1.

During feature gathering, instead of parallelizing ray samples across PEs, we parallelize channels across PEs.
Each PE is responsible for gathering one channel of all the ray samples rather than gathering all channels of one individual ray sample.
That is, each PE is dedicated to a specific bank.
For instance, in \Fig{fig:our_layout}, the four PEs are collecting the four channels of the same feature vector 3 (required by one ray sample).

\section{Hardware Support}
\label{sec:mem:hw}

\begin{figure}[t]
  \centering
  \includegraphics[width=\columnwidth]{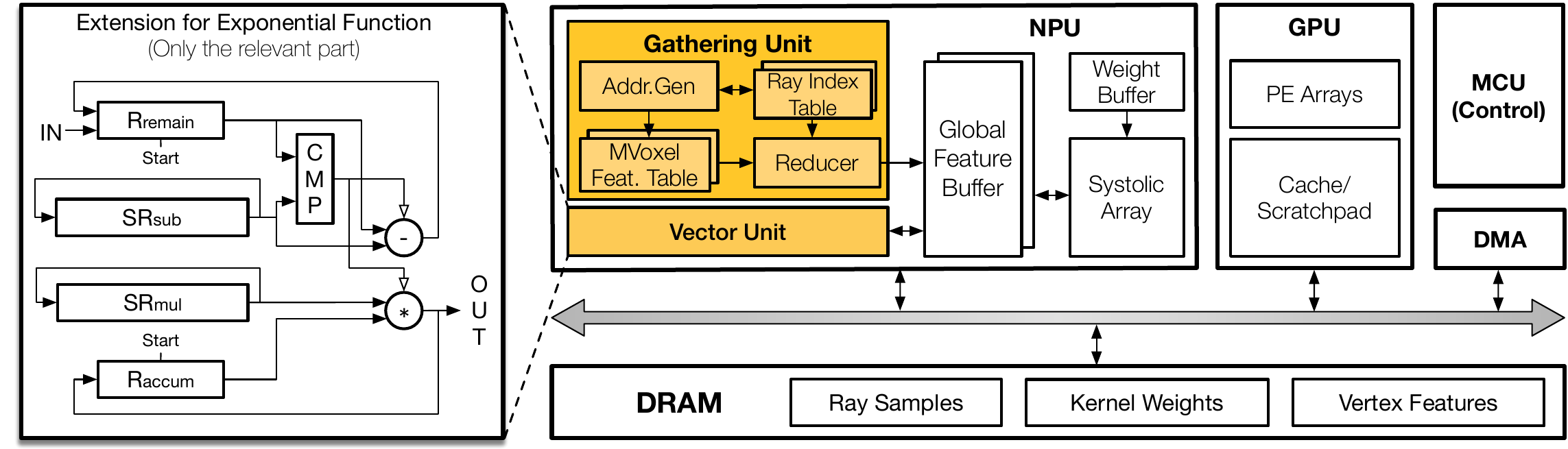}
  \caption{The SoC architecture; the uncolored is the baseline architecture; and we augment a standard systolic array-based NPU with a Gathering Unit (GU) and a vector unit; colored.
  The GU executes Feather Gathering ($\cG$) and the MAC array executes Feature Computation ($\cF$).
  GPU executes the rest. Vector Unit supports element-wise updates such as ReLU, and exponential operation.
  }
  \label{fig:arch}
\end{figure}

\begin{figure*}[t]
\centering
\includegraphics[width=\columnwidth]{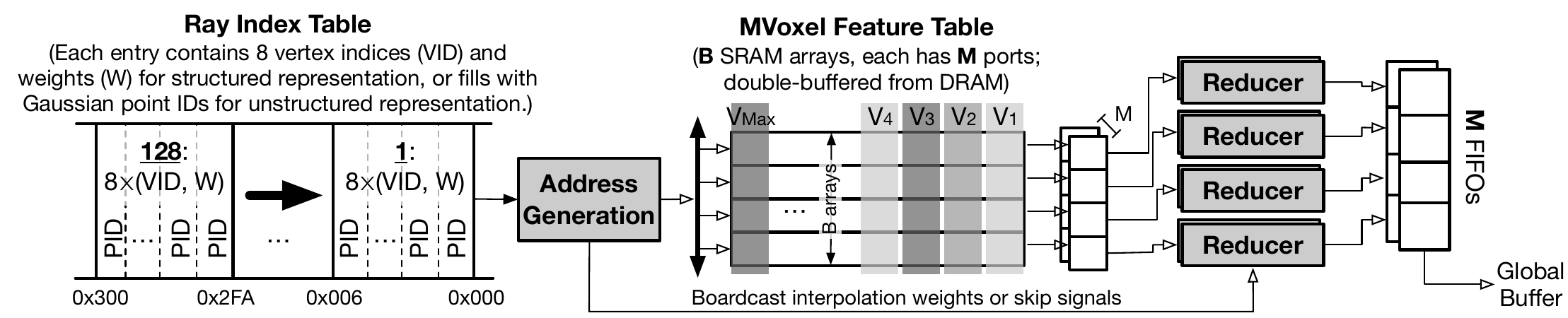}
\caption{Gathering unit.
Each RIT entry stores the information of a ray sample.
For structured representations, each entry has eight vertex IDs (VIDs) as well as the corresponding weights (W) for trilinear interpolation. For unstructured representations, each entry contains a single ID of the intersected Gaussian point (PID).
The Address Generation logic uses the VID/PIDs to access the corresponding features from the MFT, which stores the features of a MVoxel using the channel-major layout.
The VFT has $B$ individual SRAM arrays, each has $M$ ports, supporting retrieving features for $M$ ray samples simultaneously.
Trilinear interpolation is supported by $B \times M$ number of reducers to calculate the final features for $M$ ray samples.
The weights are broadcast to the reducers.
A FIFO holds the feature results before writing to the Global Feature Buffer in \Fig{fig:arch}.}
\label{fig:hw_support}
\end{figure*}


\paragraph{Architecture Overview.} We extend a standard NPU architecture to support the two memory optimizations.
\Fig{fig:arch} shows the SoC architecture, in which we augment the NPU with the new Gathering Unit (GU) and modify the vector unit to support exponential computation for 3DGS.
The baseline SoC consists of mainly a GPU and an NPU.
The GU in the NPU executes Feather Gathering ($\cG$) and the MAC array in the NPU executes Feature Computation ($\cF$).
GPU executes the rest of the computations, e.g., Ray Indexing ($\cI$) and the first three steps in \algo for the target frames. 
For 3DGS with unstructured representations, \textit{Sorting} and \textit{Spherical-Harmonics-to-Color} are also executed on GPU.



\paragraph{Vector Unit.} 
To support Gaussian splatting operation within 3DGS algorithms, we augment the vector unit to include exponential functionality. We introduce a convergence method to iteratively compute the exponential function~\cite{behrooz2000computer}. Here, we focus on explaining the computation of exponential values within the range of -1 to 1. For values beyond this interval, they still can be computed using our extension by scaling down using powers of two.

The key idea of this approach is to decompose the power $x$ into a sequence $\{ \ln(1+\alpha_{i} \times 2^{-i})\}$, where  $\alpha_{i} \in \{0, 1\}$ and $i \in [0, M)$. 
This can be done by comparing the residual value in Register $R_{\text{remain}}$ with the first element in Shift Register $SR_\text{sub}$, and enabling their subtraction (\Fig{fig:arch}). 
Meanwhile, at each iteration, the first element of Shift Register $SR_\text{sub}$ cycles to the end, while the subtraction result is written back into $R_{\text{remain}}$. 
The outcome from the comparator serves as a flag to enable the multiplication between the value in \( R_{\text{accum}} \) and the first element in Shift Register $SR_\text{mul}$ (containing \( \{(1+\alpha_{i} \times 2^{-i}), i \in [0, M)\} \)). Once the multiplication is done, the result is also written back to register $R_\text{accum}$, and the first element in Shift Register $SR_\text{mul}$ would cycle to the end. Following $M$ cycles, the final exponential value would be obtained.

\paragraph{Gathering Unit.} The GU architecture is detailed in \Fig{fig:hw_support}. This GU includes a dedicated buffer for RIT and a double-buffered MVoxel Feature Table (MFT) for streaming MVoxels from DRAM, as discussed in \Sect{sec:mem:fs}. The data layout for the MFT, optimized to eliminate bank conflicts, is outlined in \Sect{sec:mem:bank}. After the RIT entries are loaded, the address generation logic in GU processes each entry to compute the necessary addresses for retrieving the corresponding vertex or Gaussian point features from the MFT.


MFT, free of bank conflicts, is configured as $B$ individual SRAM arrays without a crossbar to reduce area overhead. 
All channels of a feature vector are accessed simultaneously, enabling a one-cycle read per vertex feature. 
For structured representations that require eight vertices per voxel, this translates to eight cycles to access all features for a ray sample. 
Conversely, unstructured representations only require a single cycle per entry. 
Each SRAM bank is designed with  $M$ ports, allowing parallel reading and processing of $M$ ray samples.


The GU uses $B \times M$ reducers to perform trilinear interpolation to calculate feature vectors for each ray sample. 
A skip-reduction flag enables bypassing trilinear interpolation for 3DGS.  
Once processed, results are stored in the Feature FIFO and then transferred to the global feature buffer in the NPU.

\paragraph{SoC Integration.}
Our hardware extensions are limited to the NPU without changing the GPU hardware.
Our design is agnostic to and, thus, integrates well with different GPU architectures --- for two reasons.  First, our hardware augmentation, i.e., the Gathering Unit, is limited to the NPU, whose communication with the GPU is dealt with by standard SoC-level interconnect (e.g., AXI) and thus accommodates different GPUs.  Second, the interaction between the NPU and the GPU is minimal in our design: the GPU simply sends the RIT through the DMA to the NPU.


\section{Experimental Setup}
\label{sec:exp}

\paragraph{Hardware Details.}
The NPU is a systolic-array-based DNN accelerator, which has a $24 \times 24$ MAC array, where each MAC unit mimics the design of that in the TPU~\cite{jouppi2017datacenter}.
Each PE consists of two 16-bit input registers, a 16-bit fixed-point MAC unit with a 32-bit accumulator register, and simple trivial control logic.
The NPU also consists of a vector unit, which can parallelize element-wise updates such as ReLU, exponential operations.
The Global Feature Buffer is configured to be double-buffered with a size of 1.5~MB at a granularity of 32~KB.
We reserve a dedicated 96~KB weight buffer to store MLP weights.

In our GU design, the RIT is double-buffered, each sized at 6~KB. Depending on the NeRF algorithms, 6~KB can store 128 entries (each accommodates eight vertices) for structured representation or 1536 Gaussian point indices for unstructured representation.
The MFT is also double-buffered, with 32 KB each (organized as $B=32$ banks each with $M=2$ ports), which can store a MVoxel ($8\times8\times8$ points) with 32 channels.
When the channel size of a feature is greater than 32, we partition the features into segments along the channel direction and load each segment sequentially.

\paragraph{Experimental Methodology.}
We directly time the GPU execution as well as the kernel launch on the mobile Volta GPU on Nvidia's Xavier SoC~\cite{xaviersoc}. The GPU power is directly measured using the built-in power sensing circuitry.
We synthesize, place, and route the datapath of 
our design 
using an EDA flow consisting of Synopsys and Cadence tools with the TSMC 16 nm FinFET technology and scale the results to 12 nm using the DeepScaleTool~\cite{stillmaker2017scaling, sarangi2021deepscaletool} so that the results can be comparable with the mobile Volta GPU on Nvidia's Xavier SoC in 12 nm node~\cite{xaviersoc}.

The SRAMs are generated using the Arm Artisan memory compiler. Power is estimated using Synopsys PrimeTimePX with annotated switching activities.
The DRAM is modeled after Micron 16 Gb LPDDR3-1600 (4 channels) according to its datasheet~\cite{micronlpddr3}. The DRAM energy is obtained using Micron System Power Calculators~\cite{microdrampower}. On average, the energy ratio between a random DRAM access and a streaming DRAM access is about 3:1, and the energy ratio between a random DRAM access and an SRAM access is about 25:1.
We build a cycle-level simulator of the architecture with the latency of each component parameterized from measurements (for GPU) and post-synthesis results of the NPU design.

\paragraph{Area Overhead.} \proj introduces minimal area overhead with GU augmentation. The major overhead is from 44~K SRAM introduced from RIT buffer and VFT buffer. The additional area overhead (0.048~mm$^2$) compared to baseline NPU is less than 2.5\%, in which the \proj-specific portion is almost negligible compared to the entire SoC area, such as 350 mm$^2$ for Nvidia Xavier~\cite{xaviersochotchips} and 108 mm$^2$ for Apple A15~\cite{applea15}. We also removed the crossbar connections in VFT buffer due to our interleaving access pattern in feature gathering. In comparison, a heavily banked SRAM with a crossbar would introduce an additional 0.036~mm$^2$ of area overhead.

\paragraph{NeRF Algorithms.} \proj can accommodate arbitrary NeRF algorithms. To demonstrate the flexibility of \proj, we evaluate three different NeRF algorithms: \textsc{Instant-NGP}~\cite{muller2022instant}, \textsc{DirectVoxGO}~\cite{sun2022direct},  \textsc{TensoRF}~\cite{chen2022tensorf}, \textsc{3DGS}~\cite{kerbl20233d} and \textsc{Compact-3DGS}~\cite{lee2023compact}, with varying model size-computation trade-offs (\Fig{fig:compute_vs_model_size}).
We evaluate algorithm quality on PSNR. 

\paragraph{Datasets.}
We use the Synthetic-NeRF dataset~\cite{mildenhall2021nerf}, which includes eight different synthetic scenes, for our evaluations. Additionally, to show a broader applicability of \proj, we use two real-world datasets, Unbounded-360~\cite{barron2022mipnerf360} (Bonsai trace) and Tanks and Temples~\cite{Knapitsch2017} (Ignatius trace), to assess \algo under real-world conditions. Although real-world datasets lack corresponding scene meshes, generating meshes from real-world images is a well-mature field (photogrammetry).

For mesh generation, we use Agisoft Metashape~\cite{metashape}, a well-established photogrammetry software. To avoid background discrepancies, we initially segment the backgrounds from all images and focus on generating meshes solely for the foreground scenes. We show that \proj achieves high accuracy even with these imperfect meshes in \Sect{sec:eval:acc}.


\paragraph{Baseline.}
Our baseline is the SoC in \Fig{fig:arch} without the GU.
It executes Ray Indexing ($\cI$) and Feature Gathering ($\cG$) on GPU and Feature Computation ($\cF$) on NPU for all frames.


\paragraph{Variants.}
We evaluate three variants of \proj to decouple the contribution proposed in our paper:
\begin{itemize}
    \item \mode{\algo}: only performs sparse radiance warping with the same hardware configuration as the baseline.
    \item \mode{\algo+FS}: same as \mode{\algo} except it includes the fully-streaming NeRF rendering.
    \item \mode{\proj}: the full version of \proj, which includes sparse radiance warping, fully-streaming NeRF rendering, and bank conflict-free interleaving (with GU support).
\end{itemize}

\paragraph{Application Scenarios.} We evaluate two application scenarios that commonly exist in AR/VR applications:
\begin{itemize}
    \item \textbf{Local Rendering}: All the computations are executed on the standalone device with the hardware described above.
    \item \textbf{Remote Rendering}: Many VR devices, such as the Oculus Quest series, can be tethered wirelessly to a remote machine (e.g., a nearby workstation or even the cloud) to accelerate rendering, whereas the local device is used for display and lightweight processing.
    How to effectively leverage the remote rendering paradigm is an active area of research, and our evaluation aims to demonstrate a particular use of remote rendering by offloading the reference frame rendering in our SPARW algorithm to a remote 2080Ti GPU via a wireless connection.
    The wireless communication energy is modeled as 100~nJ/B with a speed of 10 MB/s~\cite{liu2022augmented}.
\end{itemize}
\section{Evaluation}
\label{sec:eval}

We first demonstrate that \proj achieves quality levels comparable to the baseline (\Sect{sec:eval:acc}). 
Next, we discuss the effectiveness of \proj on real-world datasets (\Sect{sec:eval:disc}.)
We then show the speedup and energy reduction compared to the baseline hardware incorporating a dedicated DNN accelerator (\Sect{sec:eval:perf}). 
Finally, we show that \proj achieves better speedups compared to prior NeRF accelerators (\Sect{sec:eval:comp}).


\subsection{Rendering Quality}
\label{sec:eval:acc}

\begin{figure}[t]
\centering
\subfloat[Quality evaluation on Synthetic-NeRF dataset.]{
    \label{fig:synthetic_acc}	
    \includegraphics[width=0.48\columnwidth]{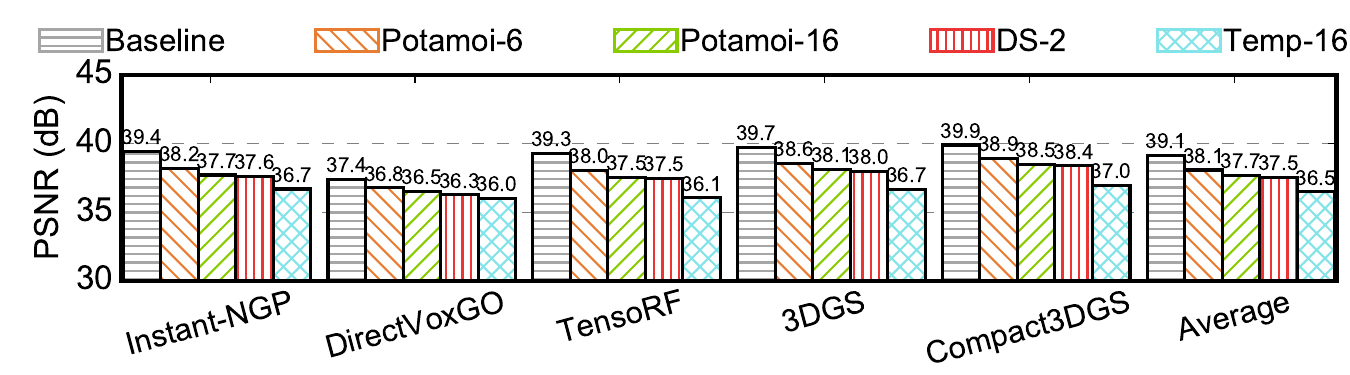} }
\hspace{-2pt}
\subfloat[Quality evaluation on real-world datasets.]{
    \label{fig:real_scene_acc}
    \includegraphics[width=0.48\columnwidth]{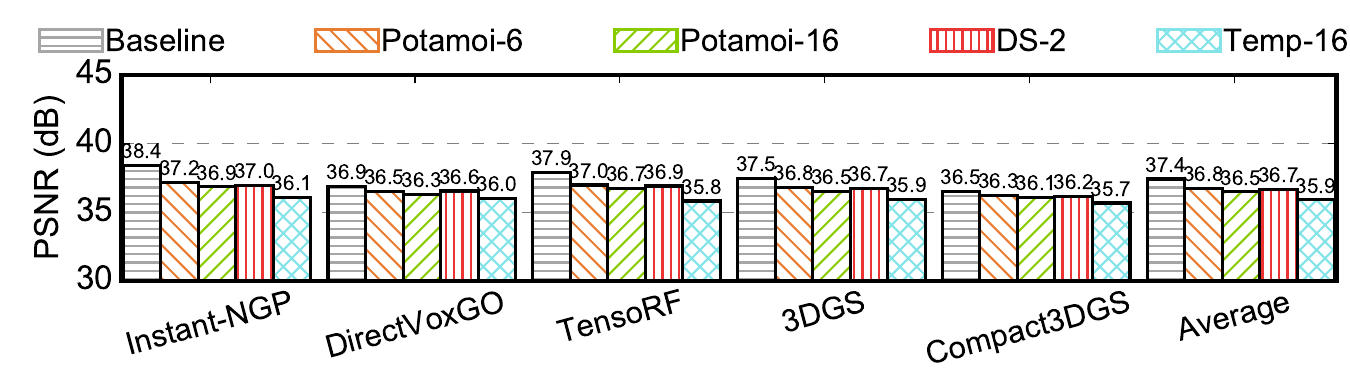} } 
\caption{Image quality comparison.
\mode{\proj-6} and \mode{\proj-16} use a warping window size (i.e., the number of target frames a reference frame is used for) of 6 and 16, respectively.
}
\label{fig:acc}
\end{figure}

\Fig{fig:acc} shows the rendering quality of applying our \algo algorithm to NeRF algorithms on both Synthetic-NeRF dataset (\Fig{fig:synthetic_acc}) and real-world scenes (\Fig{fig:real_scene_acc}). 
We use \textit{warping window} to denote the number of target frames that reuse a single reference frame. 
We consider two warping window sizes, 6 and 16. 
In addition to the baseline algorithms, we also compare against two variants, \mode{DS-2} and \mode{Temp-16}.
\mode{DS-2} first downsamples the frame by 2 for NeRF rendering and then upsamples it to the original resolution via bilinear interpolation.
\mode{Temp-16} is a method that uses a previously rendered frame as a reference frame with a warping window of 16 frames. 

On both datasets, \mode{\proj-6} retains an average PSNR drop within 1.0~dB compared to the original algorithms.
Despite \mode{\proj-16} dropping the average quality by 1.4~dB, it still has better quality compared against \mode{DS-2} and \mode{Temp-16} on Synthetic-NeRF dataset.
\mode{Temp-16} is the worst because it warps from previous frames and accumulates errors.
The quality of \mode{\proj-6} is only slightly better than \mode{DS-2} on the real-world datasets, which use a low temporal resolution (1 FPS), for which the radiance approximation does not hold well.
We will further discuss this in \Sect{sec:eval:disc}.



\subsection{Robustness}
\label{sec:eval:disc}

\begin{figure}
\begin{minipage}[t]{0.24\columnwidth}
  \centering
    \includegraphics[width=\columnwidth]{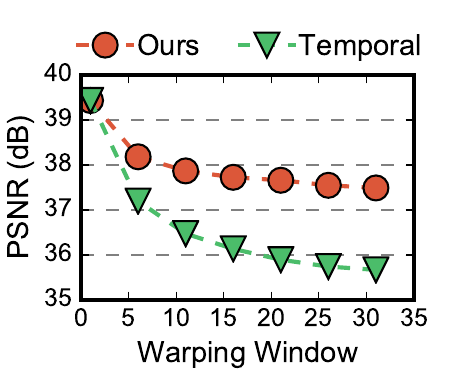}
    \caption{The accuracy sensitivity to window size between \proj and the warping-based method.}
    \label{fig:warping_cmp}
\end{minipage}
\hspace{1pt}
\begin{minipage}[t]{0.24\columnwidth}
  \centering
    \includegraphics[width=\columnwidth]{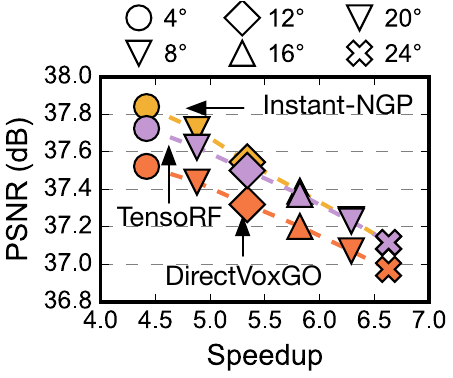}
    \caption{Speed-up and PSNR of \mode{\proj-16} under different warping thresholds $\phi$ on the 1-FPS sequence.
    }
    \label{fig:angle_threshold}
\end{minipage}
\hspace{1pt}
\begin{minipage}[t]{0.24\columnwidth}
  \centering
    \includegraphics[width=\columnwidth]{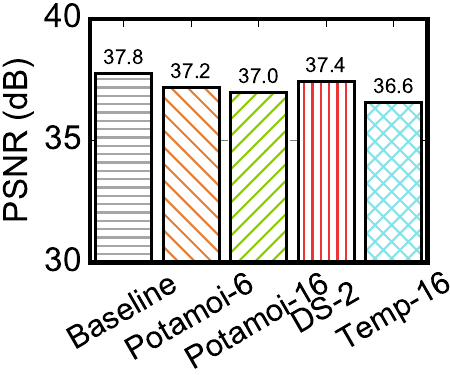}
    \caption{PSNR comparison on the Tanks and Temples dataset in sparse sequences.}
    \label{fig:sparse_acc}
\end{minipage}
\hspace{1pt}
\begin{minipage}[t]{0.24\columnwidth}
  \centering
    \includegraphics[width=\columnwidth]{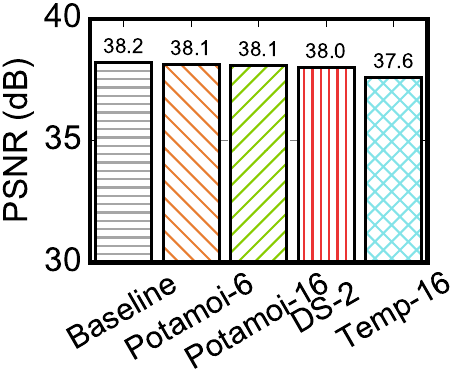}
    \caption{PSNR comparison on the Tanks and Temples dataset in dense sequences.}
    \label{fig:dense_acc}
\end{minipage}
\end{figure}

\Fig{fig:warping_cmp} shows the comparison between these two methods, \textit{reference-based warping} and \textit{temporal-based warping}. Here, we use the Synthetic-NeRF dataset and evaluate the quality under different warping window sizes using Instant-NGP~\cite{muller2022instant}. As shown in \Fig{fig:warping_cmp}, temporal-based warping drops its visual quality rapidly as the warping window size increases. At a warping window of 6, the quality of temporal-based warping is already lower than \mode{DS-2} (37.60 dB). In contrast, reference frame-based warping can still maintain competitive visual quality even with a large warping window size of 31.

We use the \textit{Ignatius} scene in the Tanks and Temples dataset to discuss the effectiveness and limitations of \mbox{\algo} on real-world scenes. 
\mbox{\Fig{fig:sparse_acc}} shows the results.
Both \mbox{\mode{\proj-6}} and \mbox{\mode{\proj-16}} have lower quality compared to \mbox{\mode{DS-2}}.
This is because the temporal resolution of the scene is extremely low (1 FPS).
Thus, consecutive frames have large differences in camera poses (i.e., $\theta$ in \mbox{\Fig{fig:intuition}} is too large), so the radiance approximation does not hold well for non-diffuse surface.


We hasten to stress that the lower quality of \mbox{\algo} here is \textit{not} fundamental to the algorithm but an artifact of the low-FPS dataset.
To evaluate \mbox{\proj} in scenarios more representative of real-time VR rendering, we use the raw video sequence from the dataset captured at 30 FPS.
The results are shown in \mbox{\Fig{fig:dense_acc}}.
In this more realistic scenario, \mbox{\mode{\proj-16}} has little quality loss over the baseline and has a similar quality compared to \mbox{\mode{DS-2}} but is about 4$\times$ faster.

While real-time VR rendering, which we target, usually has a high temporal resolution ($>$ 30 FPS), in scenarios where a dataset has low temporal resolution, our warping heuristics in \Sect{sec:algo:dd} can be used to mitigate the rendering quality loss.
\mbox{\Fig{fig:angle_threshold}} shows the speed-up and PSNR of \mbox{\mode{\proj-16}} across different warping thresholds on the challenging 1-FPS sequence of \textit{Ignatius}.
As $\phi$ reduces toward the left, the quality increases, since fewer pixels are warped and more pixels are NeRF-rendered, which also means the performance reduces.
At a threshold of 4$^{\circ}$, \mbox{\algo} has a quality drop within 0.1 dB and a speed-up of 4.3$\times$.




\subsection{Performance and Energy}
\label{sec:eval:perf}

The speedup and energy reduction are even higher when the baseline SoC uses a dedicated NPU to execute the MLPs.
To demonstrate that, we evaluate two different application scenarios: local rendering vs. remote rendering, both are common in VR.
Unless stated otherwise, we use a warping window of 16 for our evaluation.


\begin{figure}[t]
\centering
\subfloat[Speedup of local rendering.]{
    \label{fig:standalone_res_speedup}	
    \includegraphics[width=0.48\columnwidth]{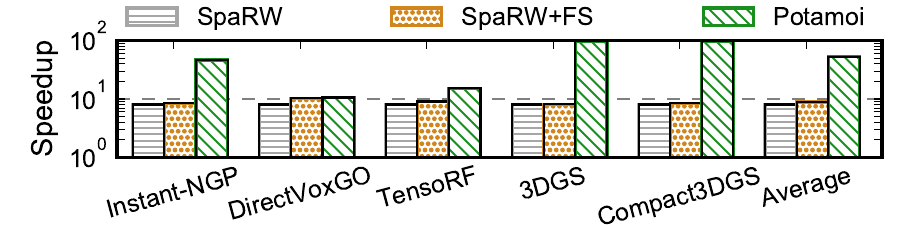}
}
\hspace{2pt}
\subfloat[Energy savings of local rendering.]{
    \label{fig:standalone_res_energy}
    \includegraphics[width=0.48\columnwidth]{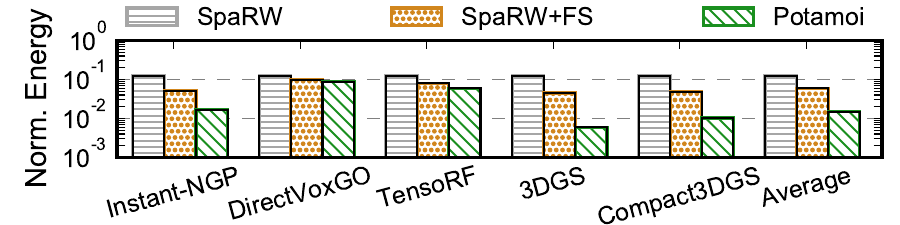}
} 
\\
\subfloat[Speedup of remote rendering.]{
    \label{fig:wireless_res_speedup}	
    \includegraphics[width=0.48\columnwidth]{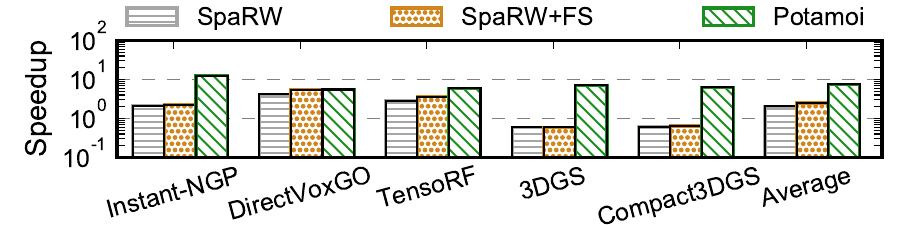}
}
\hspace{2pt}
\subfloat[Energy savings of remote rendering.]{
    \label{fig:wireless_res_energy}
    \includegraphics[width=0.48\columnwidth]{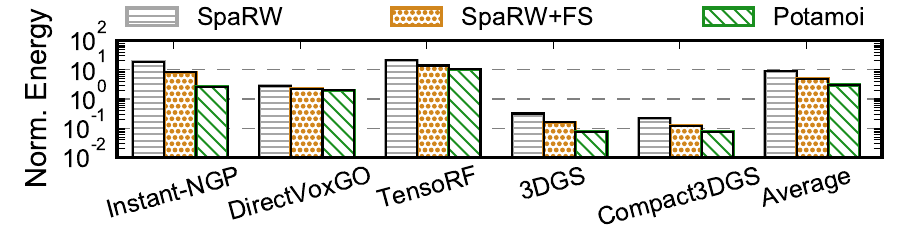}
} 
\caption{End-to-end speedup and normalized energy of our variants over the baseline with a GPU and an NPU. We evaluate two application scenarios: local rendering and remote rendering. All values are normalized to the baseline.}
\label{fig:hw_speedup_energy}
\end{figure}

\paragraph{Local Rendering.} \Fig{fig:standalone_res_speedup} and \Fig{fig:standalone_res_energy} show the speedup and normalized energy comparison in a local rendering scenario.
All results are normalized with the baseline.
On average, \mode{\algo} achieves 8.1$\times$ speedup and 8.1$\times$ energy saving on the same hardware configuration as the baseline.
With the additional assistance from fully-streaming NeRF rendering, \mode{\algo+FS} achieves an additional 1.1$\times$ speedup and 2.1$\times$ energy saving under the same hardware configuration. 

Two factors help \mode{\algo+FS} improve upon the baseline.
First, \algo algorithm reduces the amount of full-frame NeRF computation.
Second, fully-streaming NeRF rendering reduces redundant DRAM accesses. 
With GU hardware support, \mode{\proj} further boosts the speedup and energy saving to 53.1$\times$ and 67.7$\times$, respectively.
For instance, \mbox{\mode{\proj}} achieves only 9.33 FPS on Instant-NGP, whereas Instant-NGP achieves merely 0.20 FPS on the baseline hardware. On the other hand, 3DGS achieves over 1000 FPS since the baseline 3DGS is already fast.

\paragraph{Remote Rendering.}
One factor that prevents \proj from achieving higher speedup is resource contention: even though algorithmically reference frame and target frame rendering can be overlapped,
they compete for the same NPU and GPU resources.
With additional resources on a remote machine, \proj further boosts the performance. 

\Fig{fig:wireless_res_speedup} and \Fig{fig:wireless_res_energy} show the speedup and energy comparison.
In the baseline, the entire NeRF rendering executes on the remote GPU.
In our system, we map the reference frame NeRF rendering to the remote GPU and render the target frames locally.
In both cases, the remote GPU and the local device communicate the pixel data of the rendered frames.

\mode{\algo} achieves a 2.1$\times$ speedup against the baseline, while \mode{\algo+FS} achieves a 2.5$\times$ speedup by applying fully-streaming NeRF rendering.
Note that, for 3DGS and Compact-3DGS, \mode{\algo} and \mode{\algo+FS} slow down the overall performance due to the computational gap between remote GPU and local SoC.
With GU hardware support, \mode{\proj} further improves the speedup to 7.5$\times$.
Even for 3DGS and Compact-3DGS, \mode{\proj} can achieve 6.0$\times$ and 7.1$\times$ speedup.
In all cases, data communication between the remote GPU and the local device is not a bottleneck: the communication latency is $0.02\%$ of the average frame latency in \proj. 

Notably, the baseline in this scenario consumes lower energy than all three variants of \proj.
This is because when all the computations are offloaded to the remote GPU in the baseline, the main energy consumption of the local device is wireless communication.
Transferring one frame consumes almost 3.0$\times$ lower energy than rendering a frame in \proj.

\begin{figure}[t]
\centering
\begin{minipage}[t]{0.24\columnwidth}
  \centering
  \includegraphics[width=\columnwidth]{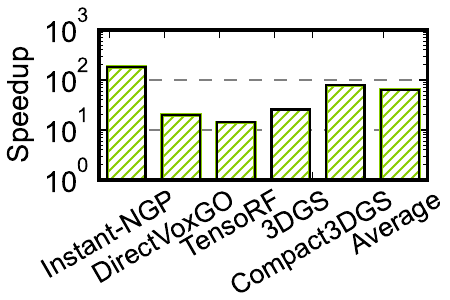}
  \caption{Speedup of feature gathering.}
  \label{fig:gu_result}
\end{minipage}
\hspace{1pt}
\begin{minipage}[t]{0.24\columnwidth}
  \centering
  \includegraphics[width=\columnwidth]{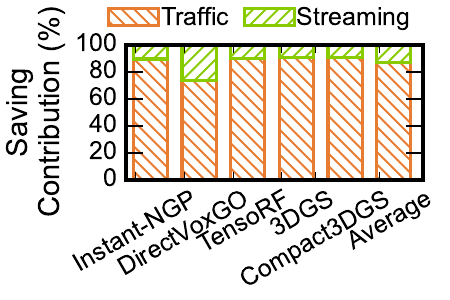}
  \caption{Memory energy saving contribution.}
  \label{fig:mem_saving_dist}
\end{minipage}
\hspace{1pt}
\begin{minipage}[t]{0.24\columnwidth}
  \centering
  \includegraphics[width=\columnwidth]{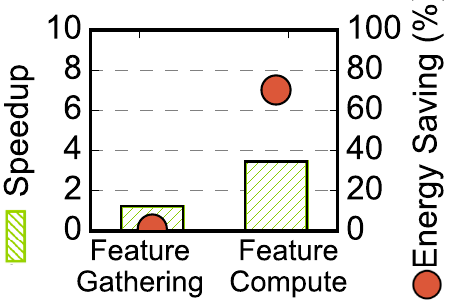}
  \caption{Energy reduction from merged Octree data structure.}
  \label{fig:octree_savings}
\end{minipage}
\hspace{1pt}
\begin{minipage}[t]{0.24\columnwidth}
  \centering
  \includegraphics[width=\columnwidth]{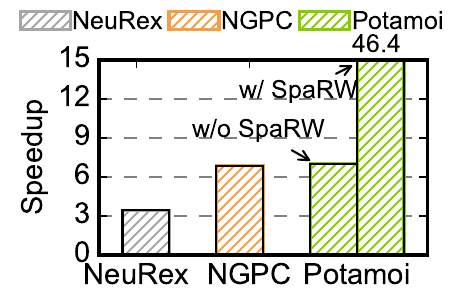}
  \caption{Performance comparison against prior works.}
  \label{fig:prior_work_comp}
\end{minipage}
\end{figure}


\paragraph{Feature Gathering ($\cG$).}
\Fig{fig:gu_result} demonstrates the speedup and energy reduction brought by GU compared to the GPU execution. 
Overall, our GU achieves  64.1$\times$ speedup while contributing to 99.9\% of the energy reduction.
This is attributed not only to our hardware acceleration of the Gather stage, but also to our data placement strategy that eliminates bank conflicts.
For instance, Instant-NGP uses hash tables which causes severe SRAM bank conflicts. \proj eliminates the irregular accesses entirely. Coupled with GU, we achieve a 182.4$\times$ speedup on Instant-NGP.

\paragraph{Memory Saving Contribution.}
Not only does \proj eliminate non-streaming DRAM access, it also reduces the overall DRAM traffic. \Fig{fig:mem_saving_dist} plots the percentage of DRAM energy reduction attributed to DRAM traffic reduction and converting random DRAM accesses to streaming accesses.
On average, 84.5\% of energy reduction is from DRAM traffic reduction.
This shows that by grouping/loading a cluster of voxels together, \proj effectively improves the reuse of each MVoxel, thus reducing the overall DRAM access. The rest of the energy reduction (15.5\%) is from converting non-streaming DRAM access to streaming DRAM access. Although reducing bank conflicts does not reduce overall energy consumption, it does improve the performance of feature gathering (\Fig{fig:gu_result}).

\paragraph{Merged Octree.} \Fig{fig:octree_savings} shows the speedup and energy savings achieved by merging the octree for unstructured representations as discussed in \Sect{sec:mem:fs}. 
Merging low-density MVoxels achieves speedup on Feature Gathering and Feature Computation stages by 1.3$\times$ and 3.5$\times$, respectively. 
This shows that merged octree improves the resource utilization of both GU and systolic array.
Regarding energy efficiency, merging octree results in a 70.1\% energy reduction of Feature Computation, which is dominated by the computation of the systolic array. Conversely, a mere 1\% of energy is saved in Feature Gathering, which is largely driven by memory operations. Note that, octree merging is performed offline, thus incurring no runtime overhead. The average memory overhead of Octree is less than 200 KB, which is much smaller compared to the overall NeRF model size.

\subsection{Compared with Prior Works}
\label{sec:eval:comp}


We also compare against two prior NeRF accelerators: \textsc{NeuRex}~\cite{lee2023neurex} and \textsc{NGPC}~\cite{mubarik2023hardware}.
Notably, both two accelerators are tailored to one particular NeRF algorithm, Instant-NGP, whereas \proj generally applies to any NeRF algorithms.

\Fig{fig:prior_work_comp} compares \mbox{\proj} with the two accelerators on Instant-NGP.
All values are normalized to the GPU baseline.
We use the data reported in the \textsc{NeuRex} paper
\footnote{The original NeuRex paper compares against Xavier NX (21 TOPS, 384 core) and our GPU baseline is Xavier (32 TOPS, 512 core).
To convert the result to the Xavier baseline, we use the actual execution time of Instant-NGP on Xavier NX and NeuRex’s speedup numbers reported in the original paper to calculate the absolute execution time of NeuRex.
Based on that, we calculate the speed-up of NeuRex over Xavier used in \Fig{fig:prior_work_comp}.}
and implement the \textsc{NGPC} architecture based on the paper's description.
For a fair comparison, we configure \mbox{\proj} to use the same amount of PEs ($24 \times 24$) as NGPC; NeuRex has a higher PE count ($32 \times 32$) as reported in the paper.
Our accelerator uses a 32 KB feature buffer, and the two baseline accelerators have larger on-chip feature buffers as described in their respective papers (i.e. 16 MB for NGPC and 64 KB for NeuRex). 


Overall, \proj without \algo algorithm demonstrates a 2.0$\times$ speedup over \textsc{NeuRex}.
The speedup against \textsc{NeuRex} is attributed to hardware augmentation of GU in \proj, which eliminates the SRAM bank conflicts in feature gathering.
By contrast, \textsc{NGPC} design inherently avoids SRAM bank conflicts (because they use one bank for all the feature vectors in one Instant-NGP level).
\proj without \algo achieves a similar speed.
However, \textsc{NGPC} requires a 16~MB on-chip buffer dedicated to storing feature encodings, which is unrealistic for a mobile SoC.
In contrast, with fully-streaming rendering algorithm, our on-chip SRAM size is only 32 KB.
With our \algo algorithm, \proj boosts the speedup to 16.4$\times$ and 8.2$\times$ against \textsc{NeuRex} and \textsc{NGPC}, respectively.

\section{Related Work}
\label{sec:related}

\paragraph{NeRF Acceleration.} NeRF rendering has drawn considerable attention in the last two years. Recent works have proposed several accelerators for NeRF algorithms~\cite{lee2023neurex, rao2022icarus, li2023instant, mubarik2023hardware, li2022rt, fu2023gen, han2023metavrain, feng2024cicero}.
However, prior designs are tailored to individual NeRF algorithms—one accelerator for one algorithm.
For example, Instant-3D~\cite{li2023instant} and \textsc{NeuRex}~\cite{lee2023neurex} accelerate the training and inference of Instant-NGP~\cite{muller2022instant}, respectively. NGPC~\cite{mubarik2023hardware} accelerates a range of neural graphic algorithms with similar hierarchical feature encodings.
ICARUS~\cite{rao2022icarus} and RT-NeRF~\cite{li2022rt}, on the other hand, design specialized architecture to speed up NeRF~\cite{mildenhall2021nerf} and TensoRF~\cite{chen2022tensorf}, respectively. Meanwhile, Gen-NeRF~\cite{fu2023gen} accelerates IBRNet~\cite{wang2021ibrnet} for novel view synthesis. 
Despite recently, Cicero~\cite{feng2024cicero} and GSCore~\cite{lee2024gscore} proposed acceleration frameworks for structured and unstructured representations, \proj proposes a fully-streaming computing framework that is generally applicable to a range of existing NeRF algorithms with minimal hardware augmentation. 


\paragraph{Memory Optimizations in Rendering.}
Our idea of full-streaming DRAM accesses is inspired by ray reordering techniques in conventional ray tracing that increase memory access locality by grouping nearby rays~\mbox{\cite{pharr1997rendering, shkurko2017dual,aila2010architecture,bikker2012improving,gribble2008coherent}}.

Our approach has three main differences.
First, we change the basic unit of reordering from rays to ray samples to accommodate the nature of NeRF.
Second, existing techniques usually manipulate rays \textit{dynamically}, since (secondary) rays are spawned at run time, which complicates the hardware design (e.g., dynamic identification of nearby rays, dynamic buffer management).
In contrast, we exploit the nature of NeRF where all ray samples are known statically and reorder ray samples only once at the beginning.
Third, ray reordering in ray tracing usually can only afford local reordering because rays are dynamically spawned, whereas we reorder ray samples \textit{globally} to guarantee fully-streaming DRAM accesses.

\paragraph{Dataflow Framework.} Dataflow optimization for deep neural network (DNN) algorithms is a well-established research area. Various techniques have been proposed to improve computational efficiency and reduce off-chip data traffic via reordering, tiling, and binding \cite{prabhakar2017plasticine, ragan2013halide, chen2018tvm}. Recently studies have aimed to fuse DNN operations to enhance data reusability and decrease data traffic across DNN layers~\cite{zheng2023tileflow, zheng2023chimera}. However, these studies focus on the dataflow scheduling of regular computations (e.g. GEMM), where optimal dataflow scheduling can be determined offline due to static data access patterns. While there are approaches designed for irregular computations~\cite{wang2021spatten, feng2022crescent, feng2020mesorasi, liu2024juno, feng2020real}, they are not suitable for NeRFs.

\proj addresses the unique challenges of NeRF rendering, where data access patterns can only be determined prior to rendering. By shifting from the conventional pixel-centric rendering to memory-centric rendering, \proj proposes a streaming framework that supports coarse-grained streaming and operation-level pipelining, effectively leveraging the unique computational characteristics of NeRF algorithms. 

\paragraph{Real-Time VR Rendering.} To achieve real-time VR rendering, prior works have leaned on image-based rendering or remote rendering~\cite{teler2001streaming, boos2016flashback, mueller2018shading, hladky2022quadstream, leng2019energy}.
Some systems directly stream rendered videos to clients, but they often suffer from bandwidth limits and require efficient video encoding~\cite{noimark2003streaming}. Other works transmit one frame that can be reused to render multiple viewpoints, but often require complex encoding methods to store texture and geometric information to address disocclusions~\cite{mueller2018shading, hladky2022quadstream}.
By leveraging the computation characteristics of NeRF, \proj proposes a straightforward yet effective approach to resolve disocclusions, achieving photo-realistic visual quality.

\section{Conclusion}
\label{sec:conc}

\proj presents an interesting idea, radiance warping, to reduce the overall pixel rendering by NeRF. This heuristic is built upon the assumption that either the object surface is non-diffuse or the angular disparity between two corresponding pairs of warping pixels is sufficiently small.
A more principal future direction is to use a lightweight DNN model to learn the material properties~\cite{gkioulekas2013inverse, azinovic2019inverse}, e.g, Bidirectional Reflectance Distribution Function (BRDF) and Bidirectional Subsurface Scattering Function Function (BSSDF)~\cite{pharr2023physically}, to approximate the warped pixel values instead.

Nevertheless, we reduce over 95\% of the MLP computation in NeRF by warping radiances computed in previous frames with less than 1 dB PSNR loss.
We also show two memory optimizations that transform the computation orders in NeRF and address the issues in both DRAM and SRAM accessing.
Collectively, we demonstrate over an order of magnitude speed-up and energy saving over a mobile Volta GPU.

\section{Acknowledgements}

We thank anonymous reviewers from TACO for their comments. The work is partially supported by the National Key R\&D Program of China under Grant 2022YFB4501400, NSFC grant (62072297 and 62222210).

\bibliographystyle{ACM-Reference-Format}
\bibliography{refs}

\end{document}